\documentclass[twocolumn]{aastex62}

\newcommand{\rprstar}{r_\mathrm{P}/R_*}

\received{January 1, 2018}
\revised{January 7, 2018}
\accepted{\today}
\submitjournal{the Astronomical Journal}

\shorttitle{The young planetary system K2-25}
\shortauthors{Kain et al.}

\usepackage{amsmath}
\usepackage{mathtools}
\usepackage{booktabs}
\usepackage{xcolor}

\begin{document}

\title{The young planetary system K2-25: constraints on companions and starspots}

\correspondingauthor{Isabel J. Kain}
\email{kain.i@husky.neu.edu, ijkain@gmail.com}

\author[0000-0001-9894-5229]{Isabel J. Kain}
\affiliation{Department of Physics, Northeastern University, Boston, MA 02115, USA}
\affiliation{Kavli Institute for Astrophysics and Space Research, Massachusetts Institute of Technology, Cambridge, MA 02139, USA}
\nocollaboration

\author[0000-0003-4150-841X]{Elisabeth R. Newton}
\affiliation{Department of Physics and Astronomy, Dartmouth College, Hanover, NH 03755, USA}
\affiliation{Kavli Institute for Astrophysics and Space Research, Massachusetts Institute of Technology, Cambridge, MA 02139, USA}

\author[0000-0001-7730-2240]{Jason A. Dittmann}
\affiliation{Kavli Institute for Astrophysics and Space Research, Massachusetts Institute of Technology, Cambridge, MA 02139, USA}

\author{Jonathan M. Irwin}
\affiliation{Harvard-Smithsonian Center for Astrophysics, Cambridge, MA 02138, USA}

\author[0000-0003-3654-1602]{Andrew W. Mann}
\affiliation{Department of Physics and Astronomy, University of North Carolina at Chapel Hill, Chapel Hill, NC 27599, USA}

\author[0000-0001-5729-6576]{Pa Chia Thao}
\affiliation{Department of Physics and Astronomy, University of North Carolina at Chapel Hill, Chapel Hill, NC 27599, USA}

\author[0000-0002-9003-484X]{David Charbonneau}
\affiliation{Harvard-Smithsonian Center for Astrophysics, Cambridge, MA 02138, USA}

\author[0000-0001-6031-9513]{Jennifer G. Winters}
\affiliation{Harvard-Smithsonian Center for Astrophysics, Cambridge, MA 02138, USA}



\begin{abstract}

The abundance of planets with orbital periods of a few to tens of days suggests that exoplanets experience complex dynamical histories. Planets in young stellar clusters or associations have well-constrained ages and therefore provide an opportunity to explore the dynamical evolution of exoplanets. K2-25b is a Neptune-sized planet in an eccentric, 3.48 day orbit around an M4.5 dwarf star in the Hyades cluster (650 Myr). In order to investigate its non-zero eccentricity and tight orbit, we analyze transit timing variations (TTVs) which could reveal clues to the migration processes that may have acted on the planet. We obtain 12 non-consecutive transits using the MEarth Observatories and long-term photometric monitoring, which we combine with 10 transits from the \textit{Spitzer} Space Telescope and 20 transits from {\it K2}. Tables of MEarth photometry accompany this work. We fit each transit lightcurve independently. We first investigate whether inhomogeneities on the stellar surface (such as spots or plages) are differentially affecting our transit observations. The measured transit depth does not vary significantly between transits, though we see some deviations from the fiducial transit model. We then looked for TTVs as evidence of a non-transiting perturber in the system. We find no evidence for $>1$ $M_\oplus$ mass companions within a 2:1 period ratio, or for $>5$ $M_\oplus$ mass planets within a 7:2 period ratio.


\end{abstract}



\section{Introduction} \label{sec:intro}

Planetary migration is thought to play an important role in shaping the final architectures of planetary systems. For example, migration results in the highly visible population of hot Jupiters \citep{Dawson2018OriginsJupiters} and also may explain the systems of tightly packed sub Neptunes \citep{Ford2014ArchitecturesFormation.}. 

The population of short-period planets includes a subset similar in radius to Neptune, sometimes referred to as hot Neptunes. There has been some speculation that hot Neptunes are part of a broader planet population that also includes hot Jupiters: perhaps the smaller close-in Neptunes are hot Jupiters that have been stripped of their atmospheres, or perhaps these two classes of planets share migration histories. Although mass loss rates of $~10^{11}-10^{12}$ g/s have been detected in Hot Jupiters, the breakdown of energy-limited escape at high-incident fluxes suggests that Neptune-size planets are not the remnants of close-in Jupiters \citep{Murray-Clay2008ATMOSPHERICJUPITERS}. \citet{Dong2018LAMOSTElements} suggest common formation or migration histories, finding that hot Neptunes and hot Jupiters share a similar preference for singly-transiting systems and metal-rich host stars.

Young exoplanets, which may be in the midst of their most dynamically active years, can provide unique insight into the origins of the planetary systems observed around typical field-aged stars (ages $\gtrsim1$ Gyr). Our focus in this paper is the young exoplanet K2-25b \citep{Mann2016ZODIACALCLUSTER}, an eccentric ($e=0.27$) short-period ($P=3.48$ d) hot Neptune in a close orbit ($a=0.035$ au) around a mid M dwarf ($M_*=0.26 M_\odot$, $R_*=0.29 R_\odot$). K2-25 is a member of the Hyades cluster, the age of which is estimated to be approximately 650 Myr \citep{Perryman1998TheAge, DeGennaro2009INVERTINGHYADES, 2018ApJ...856...40M, 2018ApJ...863...67G}. We note that \citet{2015ApJ...807...24B} and \citet{2018A&A...616A..10G} have suggested an $800$ Myr age for the Hyades \citep[see][for further discussion]{2018ApJ...863...67G}.

\citet{Mann2016ZODIACALCLUSTER} and \citep{David2016AStar} independently identified K2-25b using data from the K2 ecliptic plane survey. Though the sampling of K2 data is such that this short transit is poorly resolved in the 30-minute cadence K2 data, \citet{Mann2016ZODIACALCLUSTER} were able to validate the planet using high contrast imaging, radial velocity measurements, and statistical arguments. The short duration of the transit and constraints on the stellar density suggested that K2-25b is on an eccentric orbit, an effect which is sometimes called the ``photoeccentric effect'' \citep[e.g.][]{Barnes2007EffectsCurves, Ford2008CharacterizingObservations, Dawson2012ThePlanets}. In a companion paper, \citet{ThaoInPrep} update the stellar parameters using the new {\it Gaia} parallax, and report a moderate eccentricity for K2-25b via this effect by simultaneously modeling all observed transits (Table \ref{tab:params}). Given the star's youth and the clear $1.88$ day photometric rotational modulation seen in the {\it K2} discovery lightcurve, we note the possibility of starspots impacting the transit parameters.

K2-25b may represent a precursor of the hot Neptunes found around older stars. A notable example is Gl 436b \citep{Butler2004A436}, which is also a short-period Neptune-sized planet with a non-zero eccentricity. The origin of Gl 436b's high eccentricity remains unclear, though both Kozai migration spurred by a distant companion \citep{Bourrier2017OrbitalStar} and continuous excitation by a planetary companion that is not in a near mean motion resonance \citep{Batygin2009AEccentricity} have been explored as possible explanations for these characteristics. However, no exterior companion has been detected to date.
In contrast, the close-orbiting Neptune mass planet HD 219828b \citep{Melo2007AHD219828} has an eccentricity consistent with zero, but the system includes an outer, highly eccentric planet or brown dwarf that likely influenced its dynamical evolution \citep{Santos2016AnStar}.  

We present new transit observations and long-term photometric monitoring of K2-25b obtained using the MEarth observatories in order to probe the impact of starspots on our transit data and to look for non-transiting companions via transit timing variations \citep[TTVs;][]{Holman2004TheEarth, Agol2004OnTransits}. We combine our data with transits observed from the \textit{Spitzer} and \textit{K2} space telescopes (Section \ref{sec:data}). We detail our analysis method, in which we fit each transit independently (Section \ref{sec:fitting}). We test for the impact of stellar surface inhomogeneities on the measured transit properties (Section \ref{sec:spots}) and look for evidence of companions using TTVs (Section \ref{sec:companions}). We conclude by discussing K2-25b in the context of planet migration scenarios (Section \ref{sec:discussion}) and presenting a brief summary of our work (Section \ref{sec:summary}).

Throughout this work, we will denote individual transits by the transit number, counting up from the transit ephemeris given in \citet{Mann2016ZODIACALCLUSTER}.


\section{Data} \label{sec:data}

\begin{figure*}[t]
\centering
\includegraphics[width=0.24\textwidth]{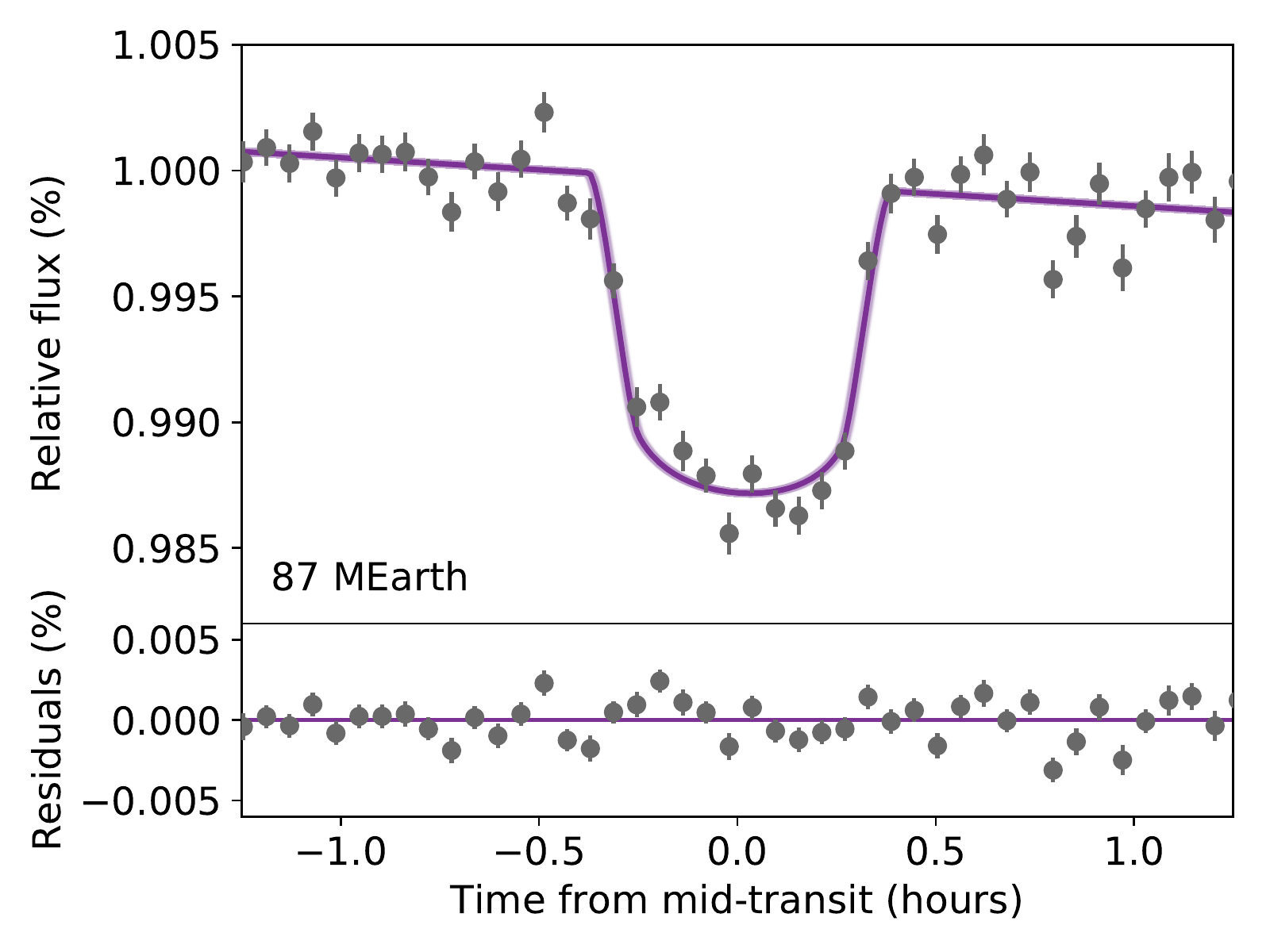}
\includegraphics[width=0.24\textwidth]{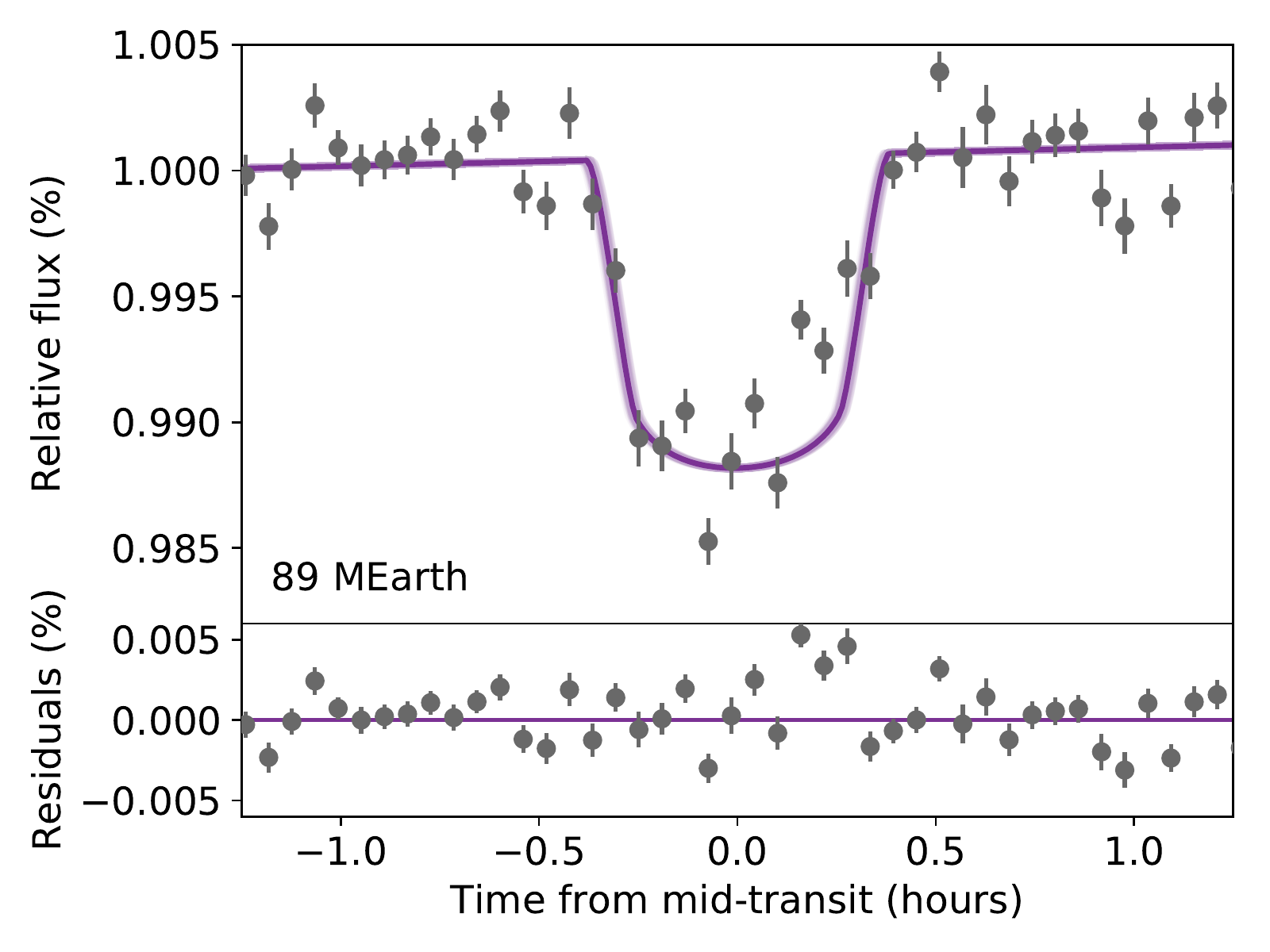}
\includegraphics[width=0.24\textwidth]{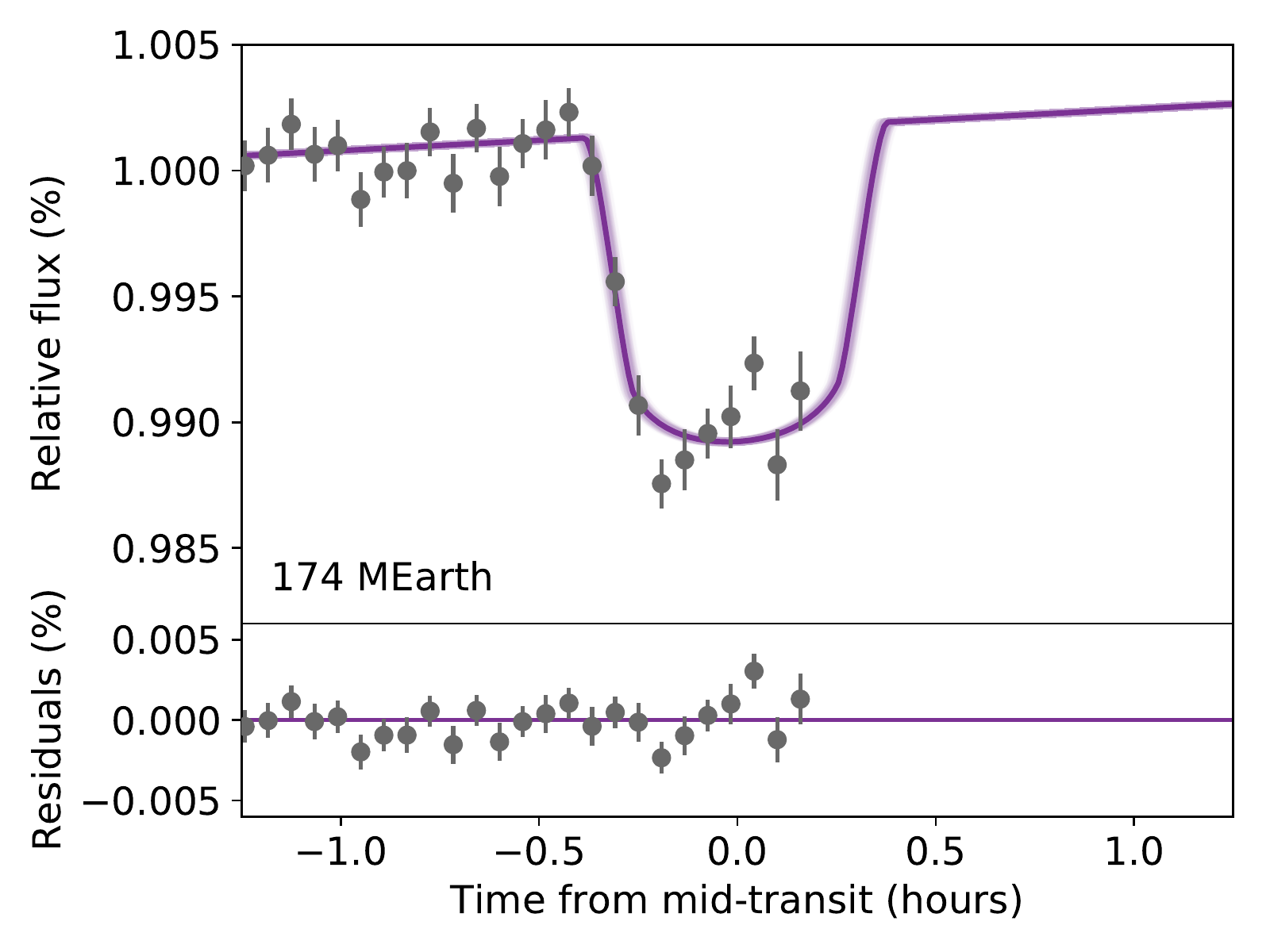}
\includegraphics[width=0.24\textwidth]{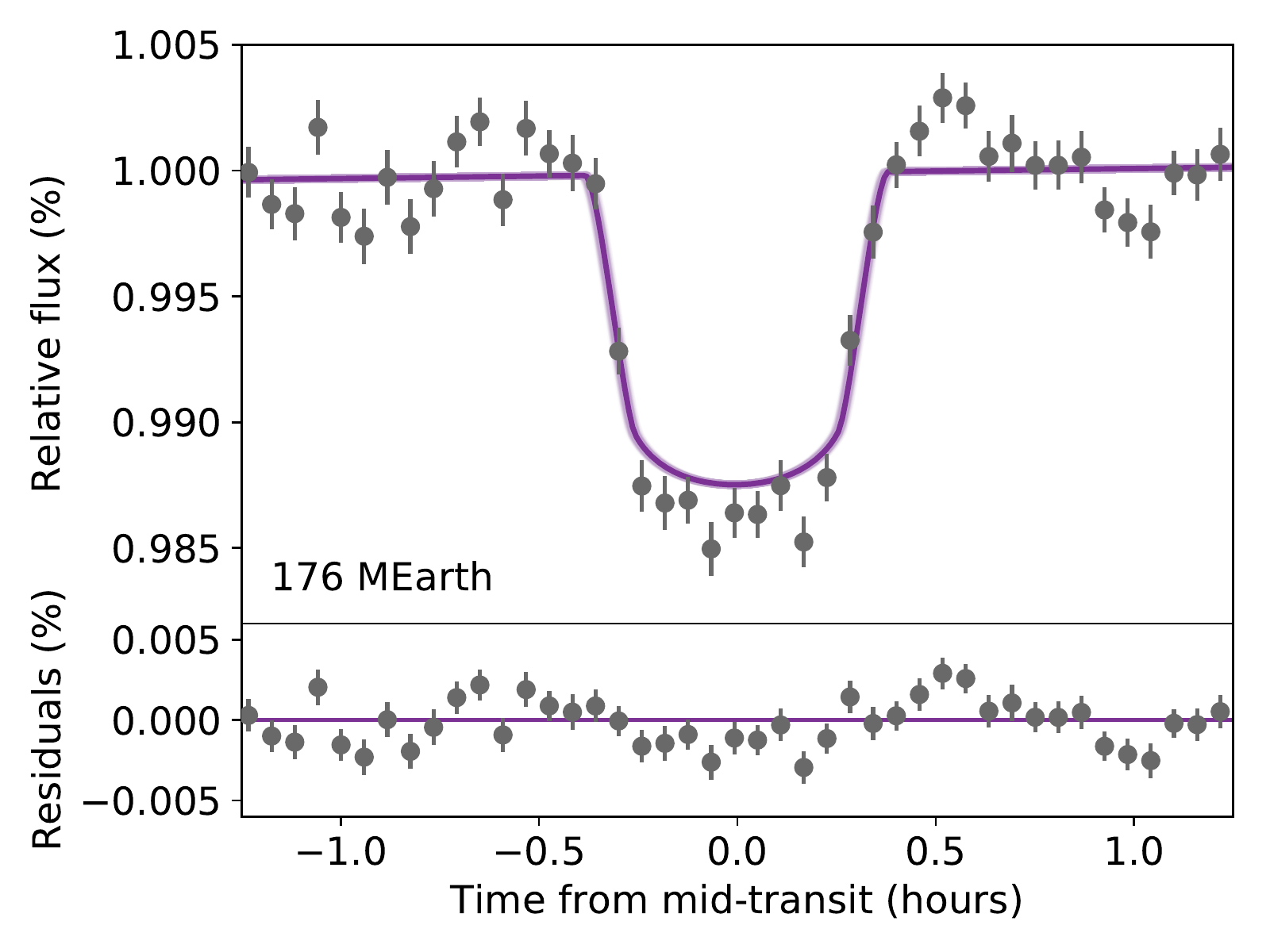}
\includegraphics[width=0.24\textwidth]{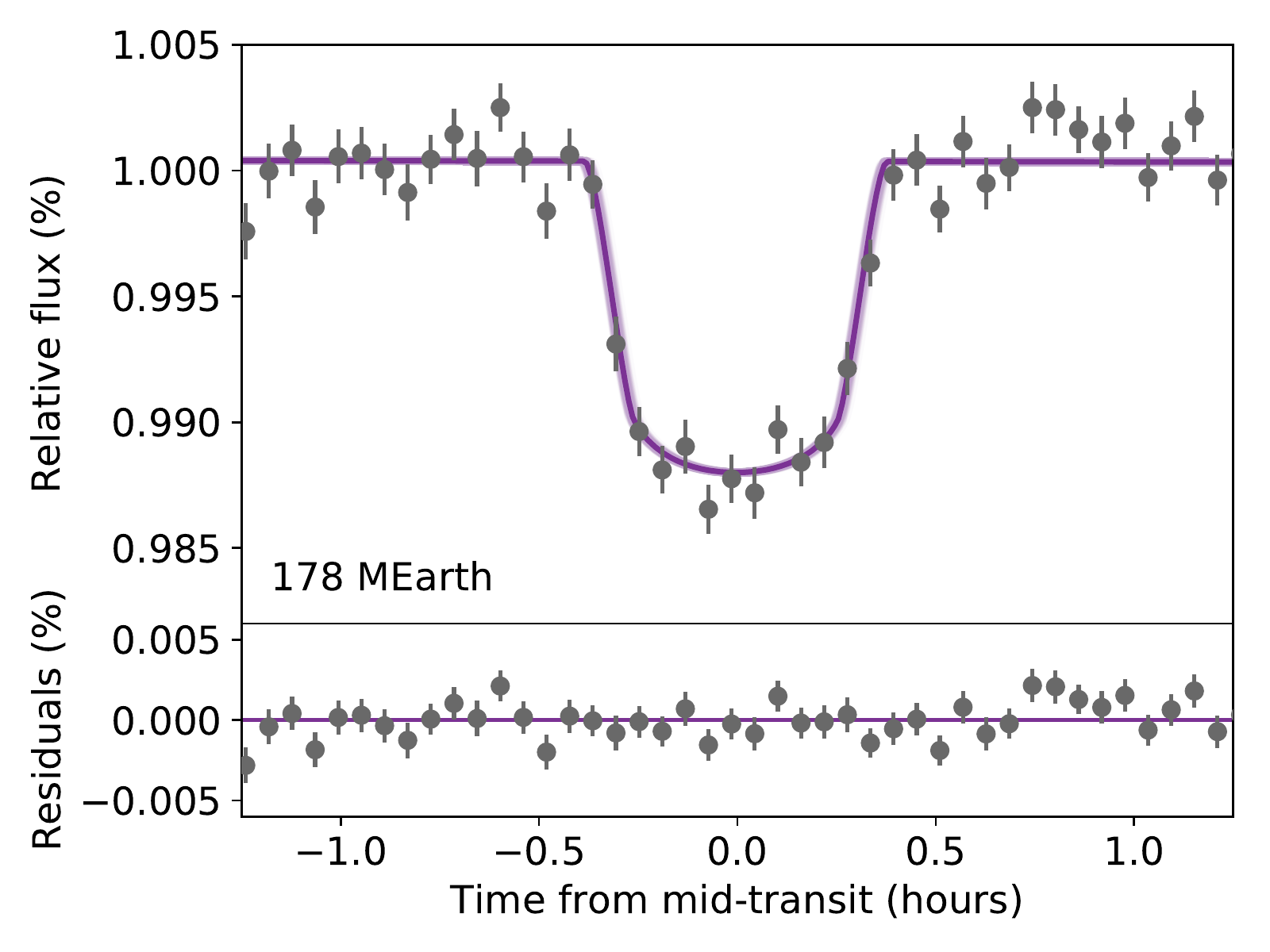}
\includegraphics[width=0.24\textwidth]{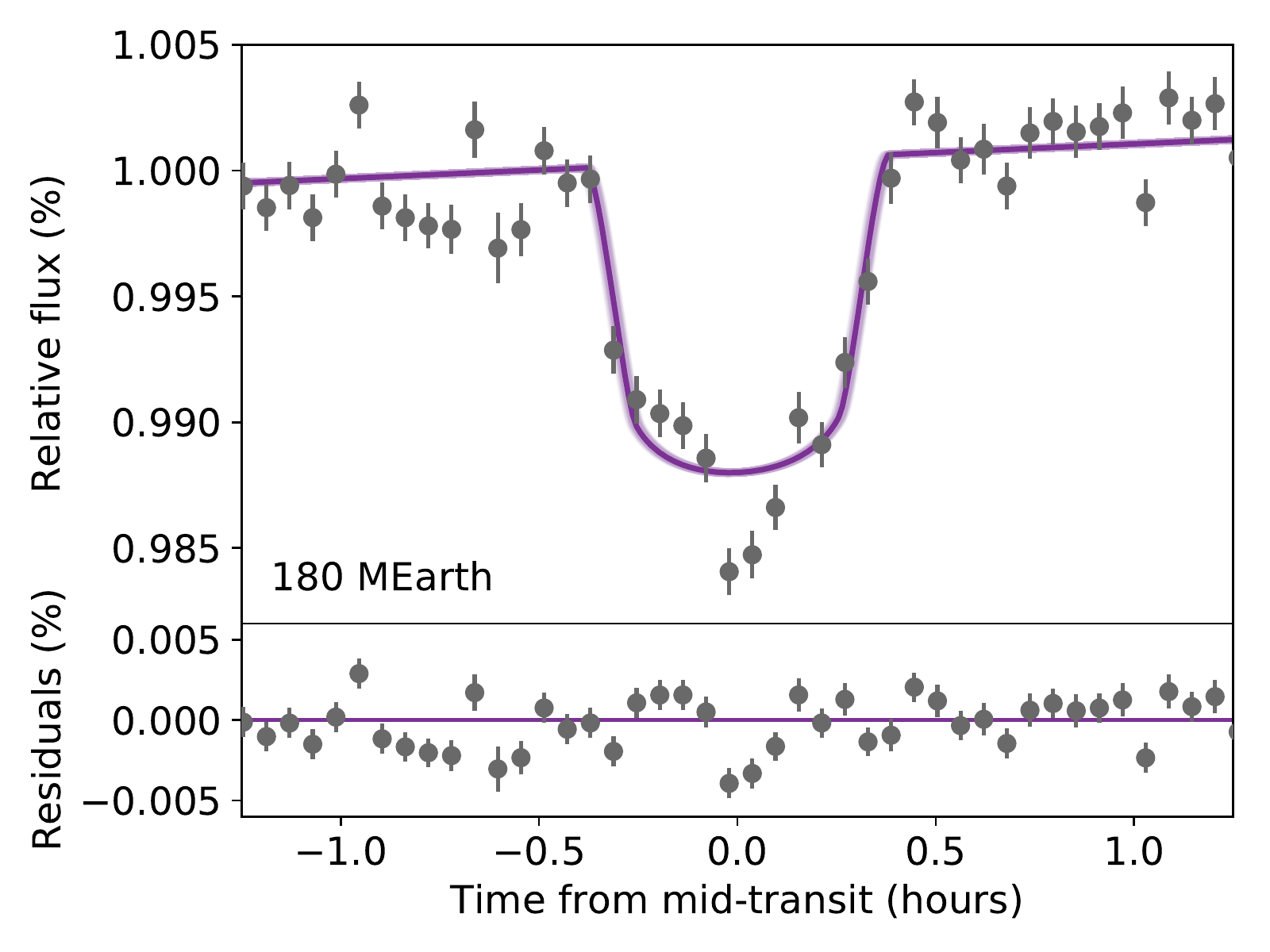}
\includegraphics[width=0.24\textwidth]{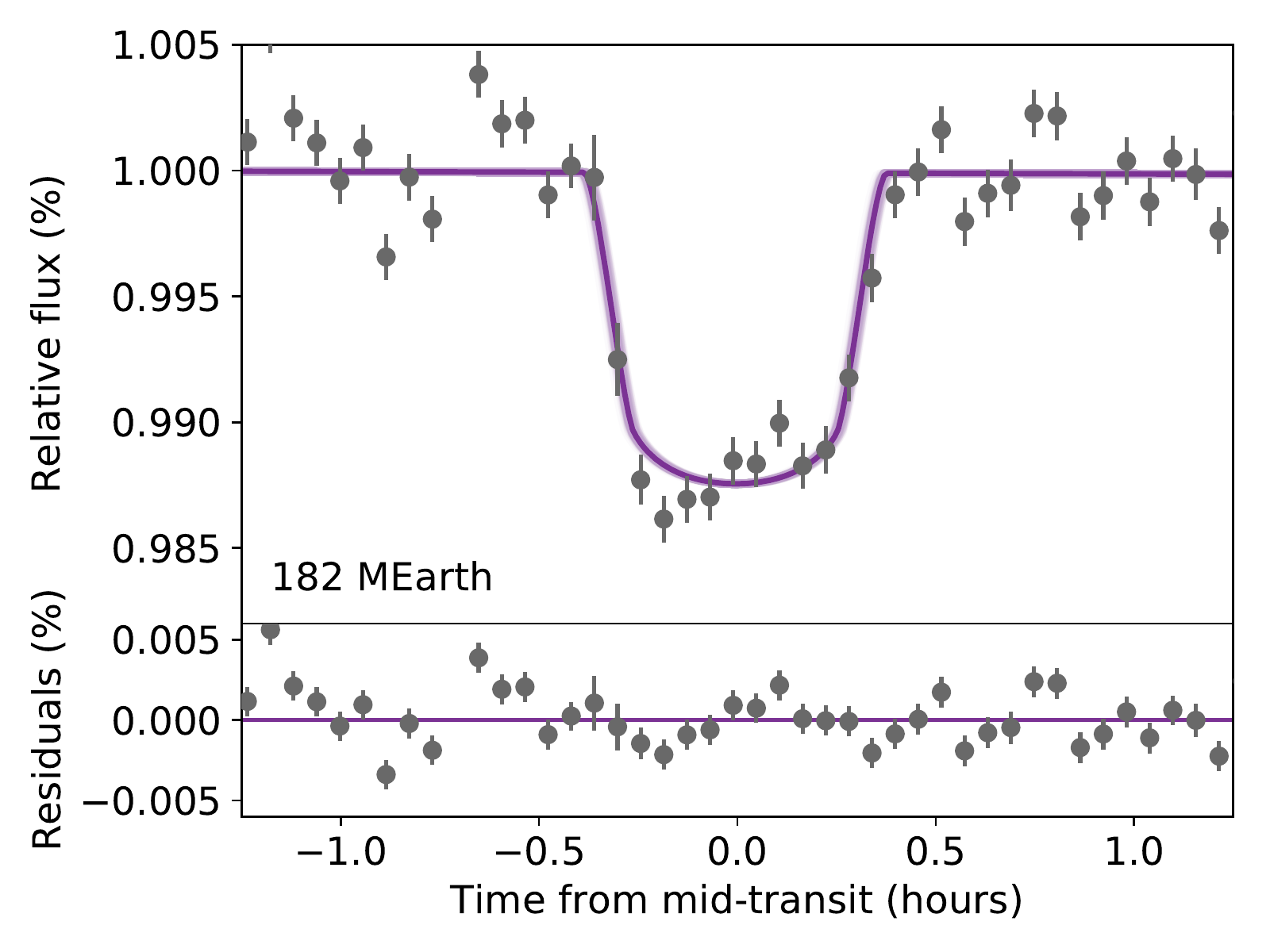}
\includegraphics[width=0.24\textwidth]{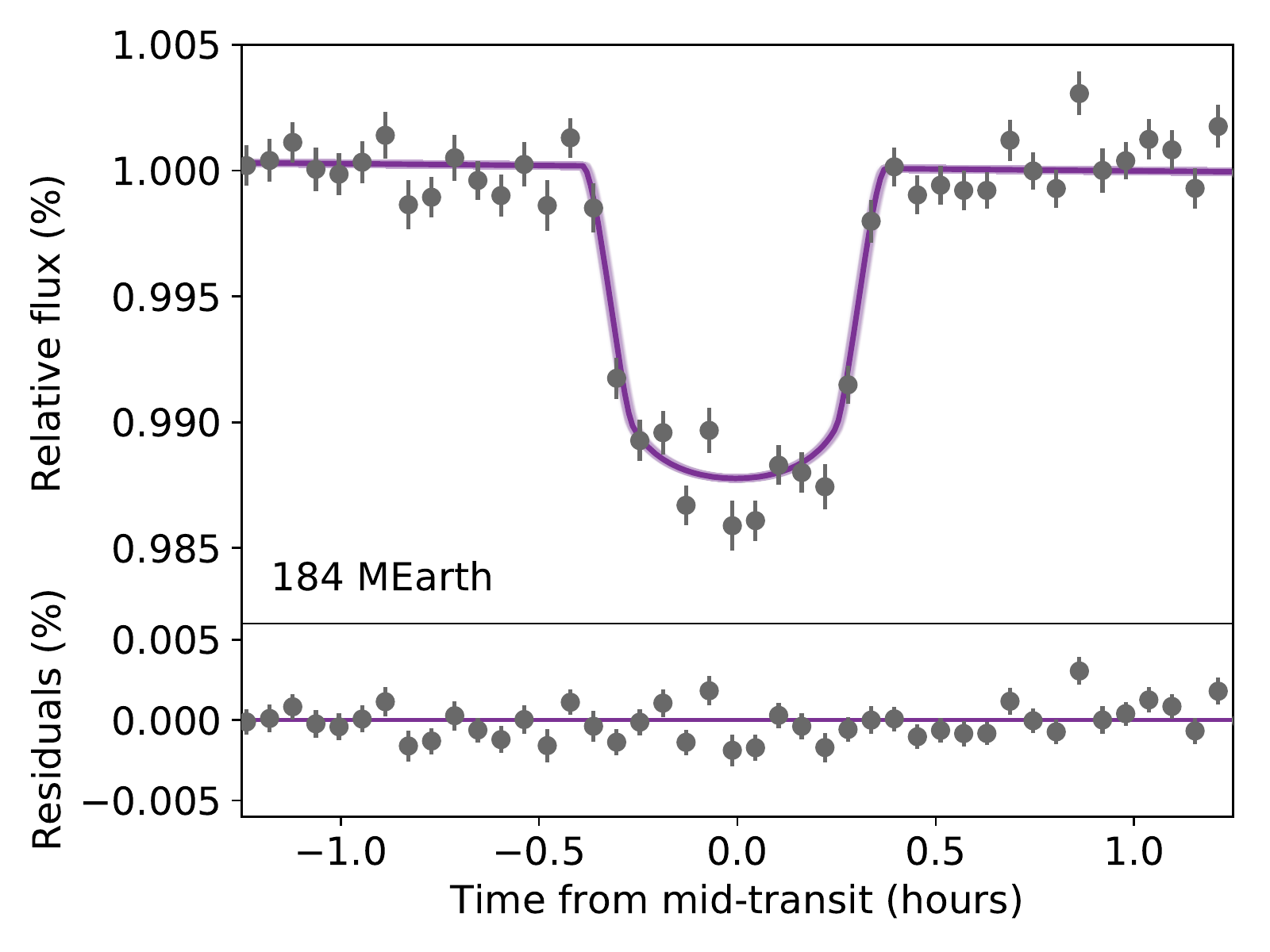}
\includegraphics[width=0.24\textwidth]{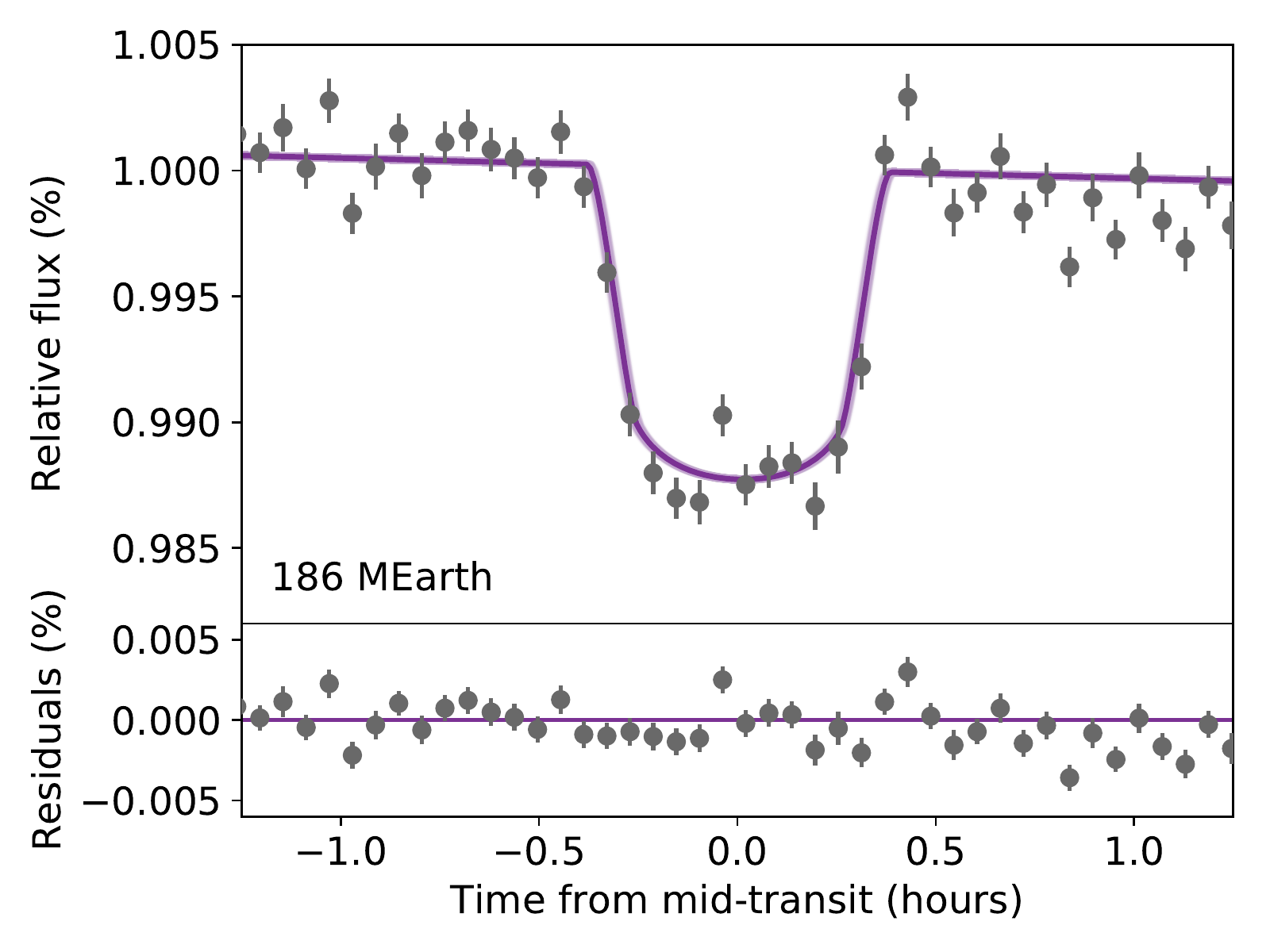}
\includegraphics[width=0.24\textwidth]{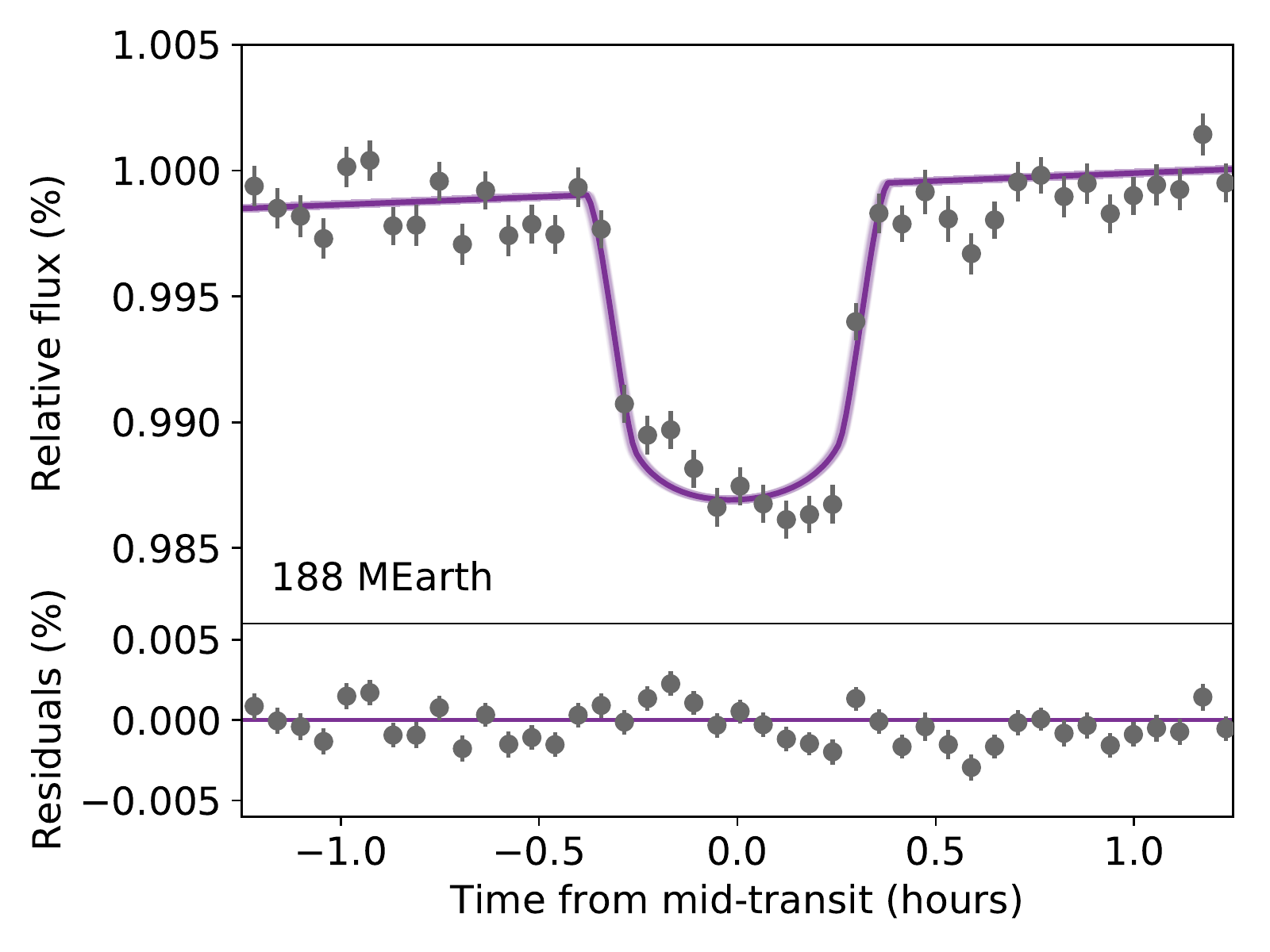}
\includegraphics[width=0.24\textwidth]{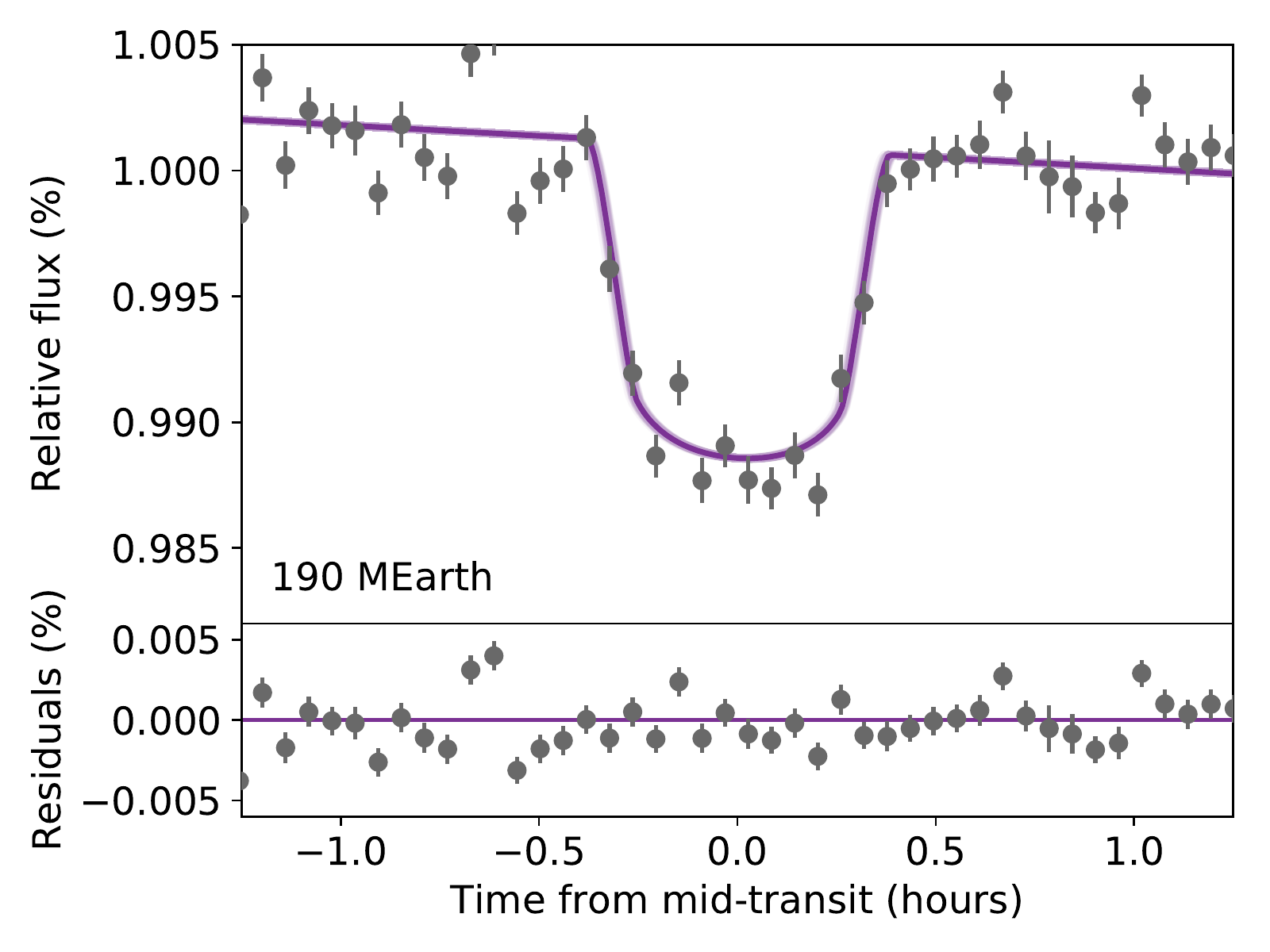}
\includegraphics[width=0.24\textwidth]{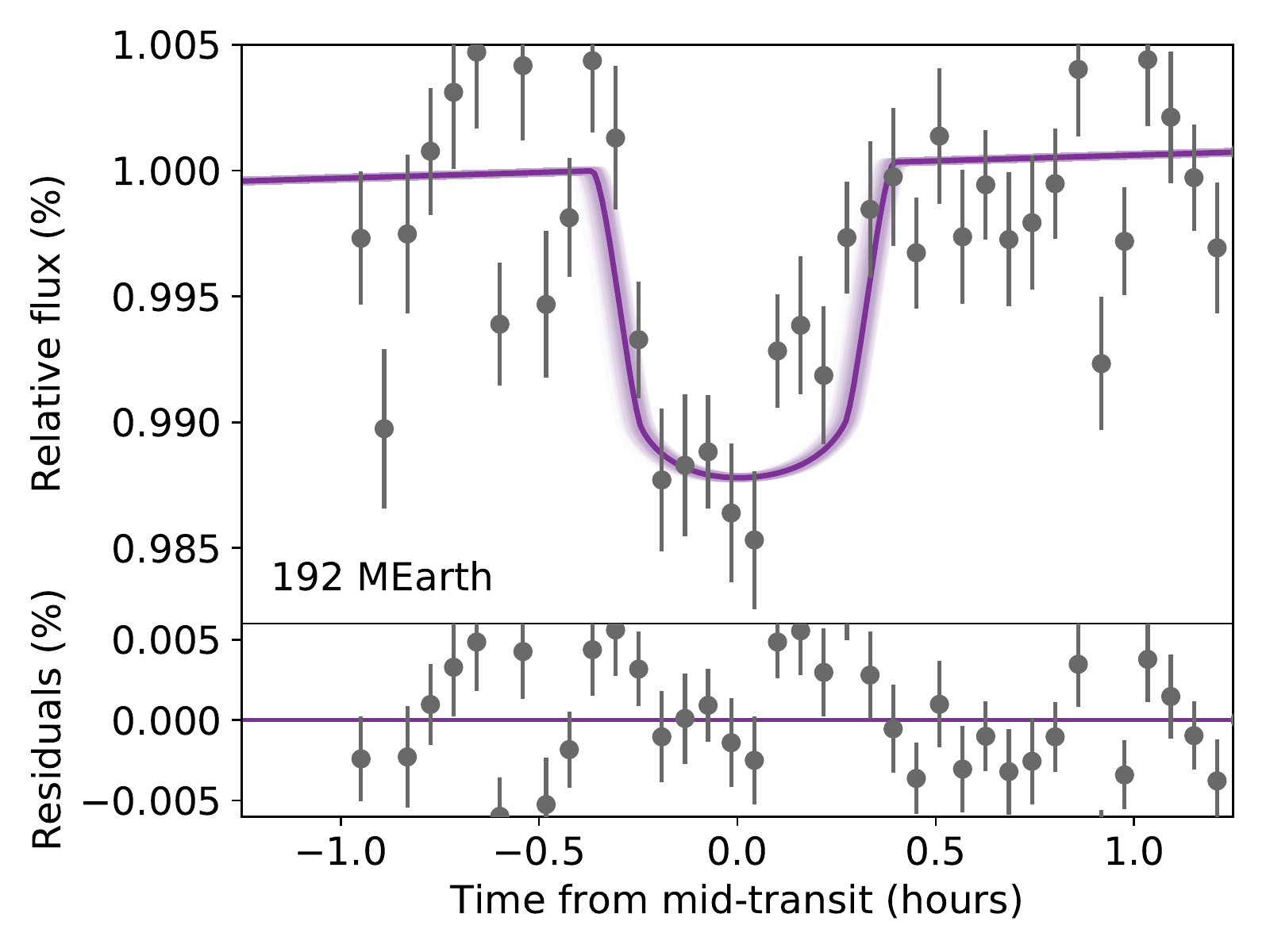}
\includegraphics[width=0.24\textwidth]{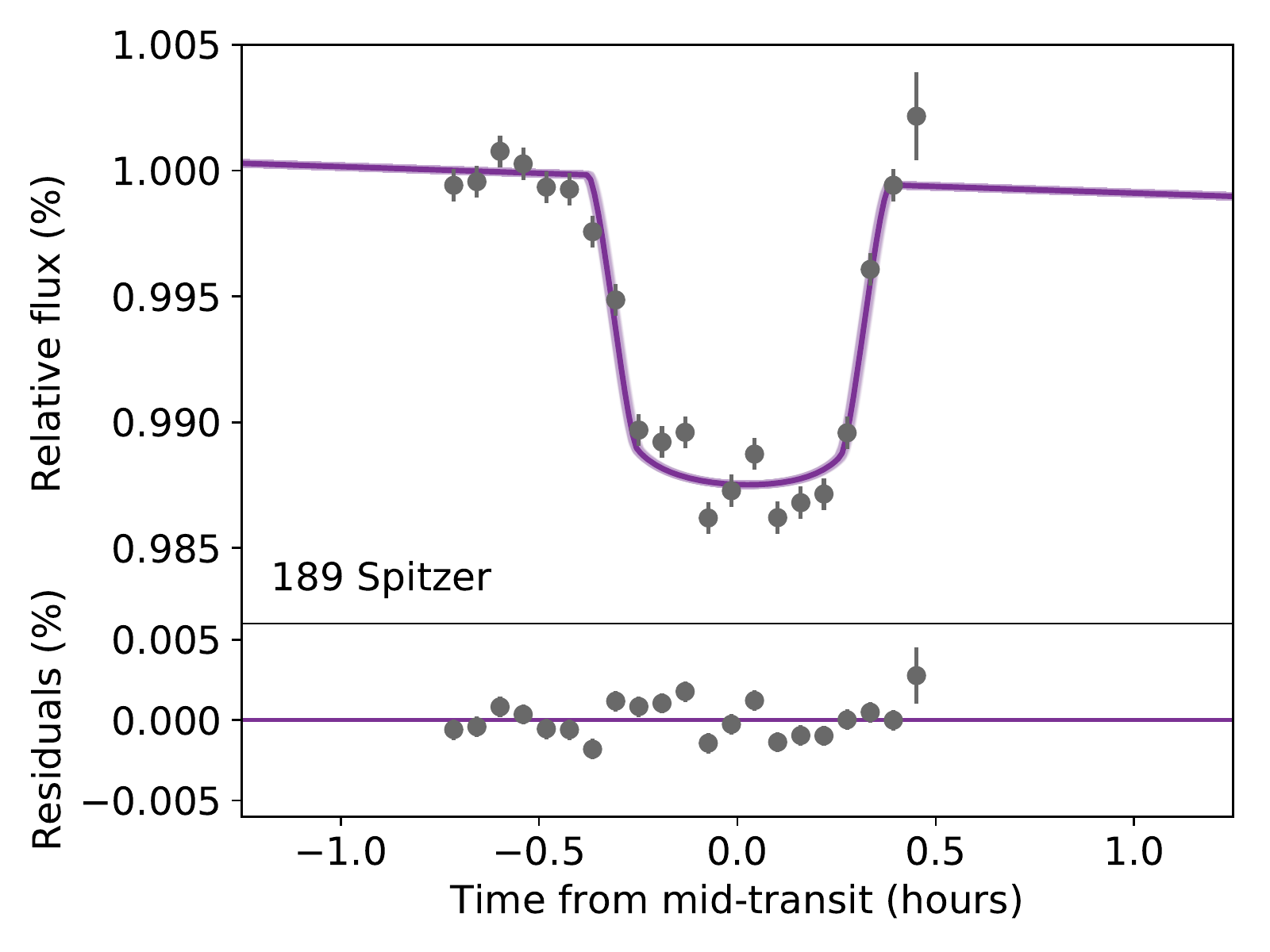}
\includegraphics[width=0.24\textwidth]{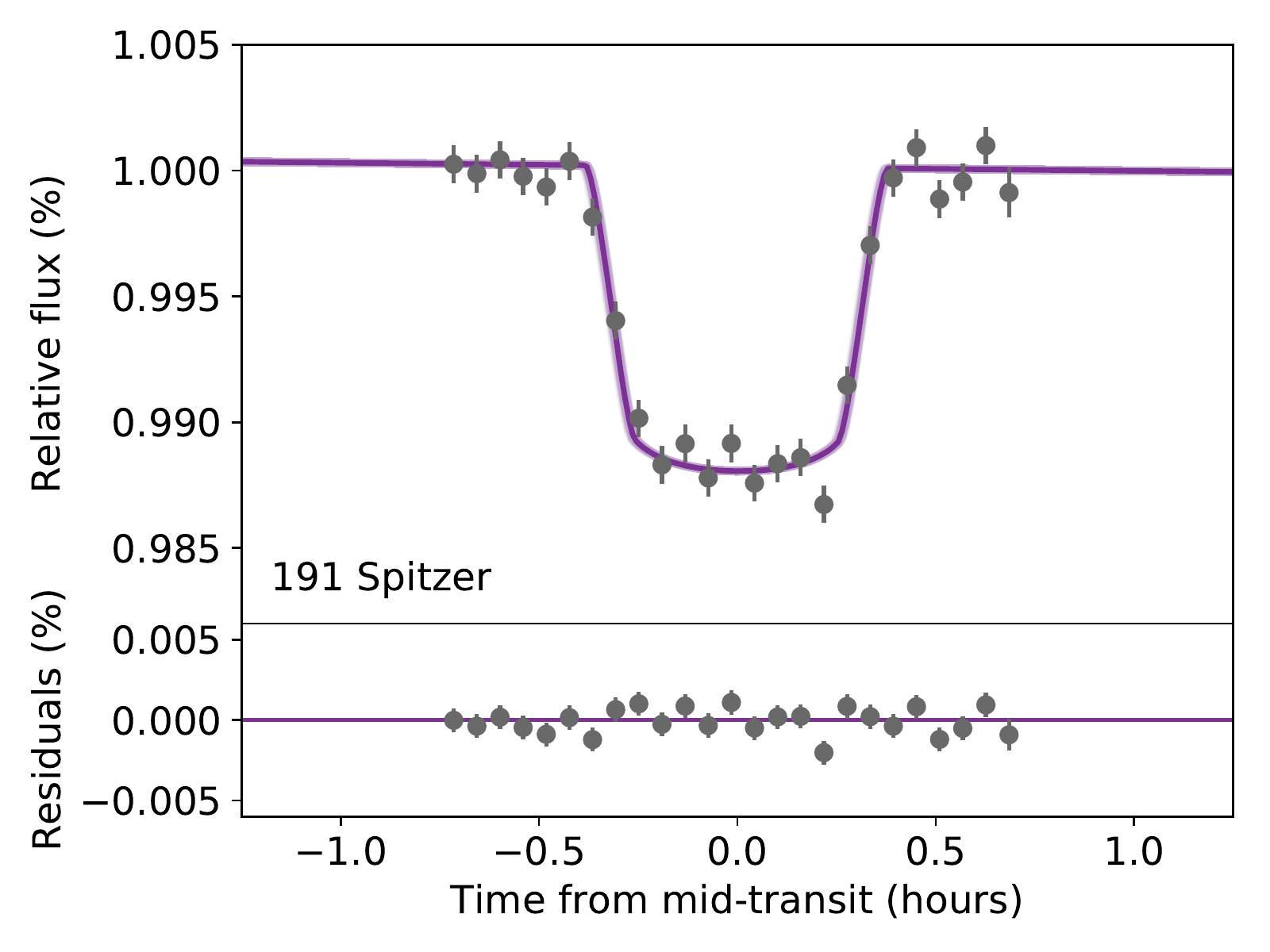}	
\includegraphics[width=0.24\textwidth]{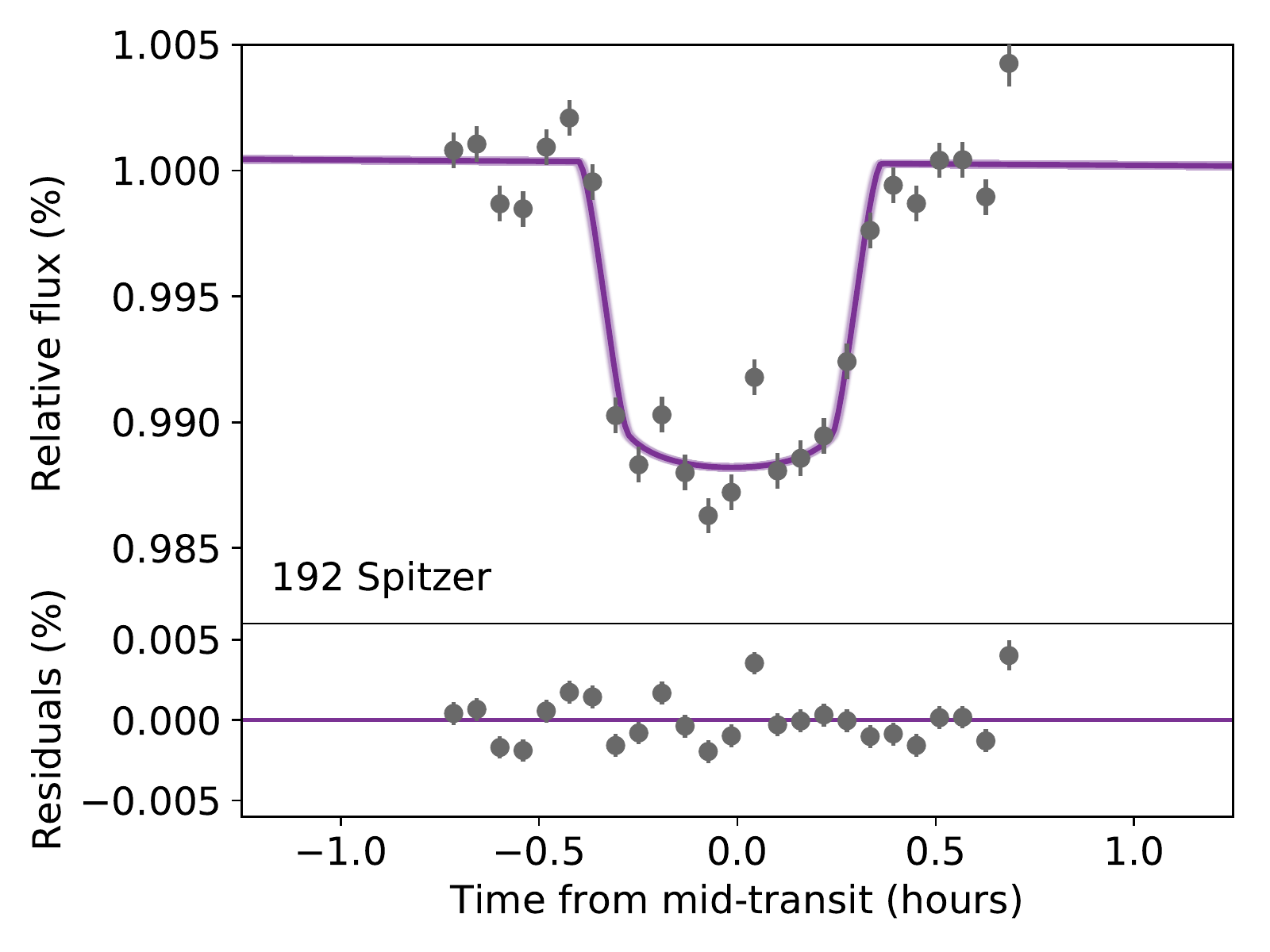}	
\includegraphics[width=0.24\textwidth]{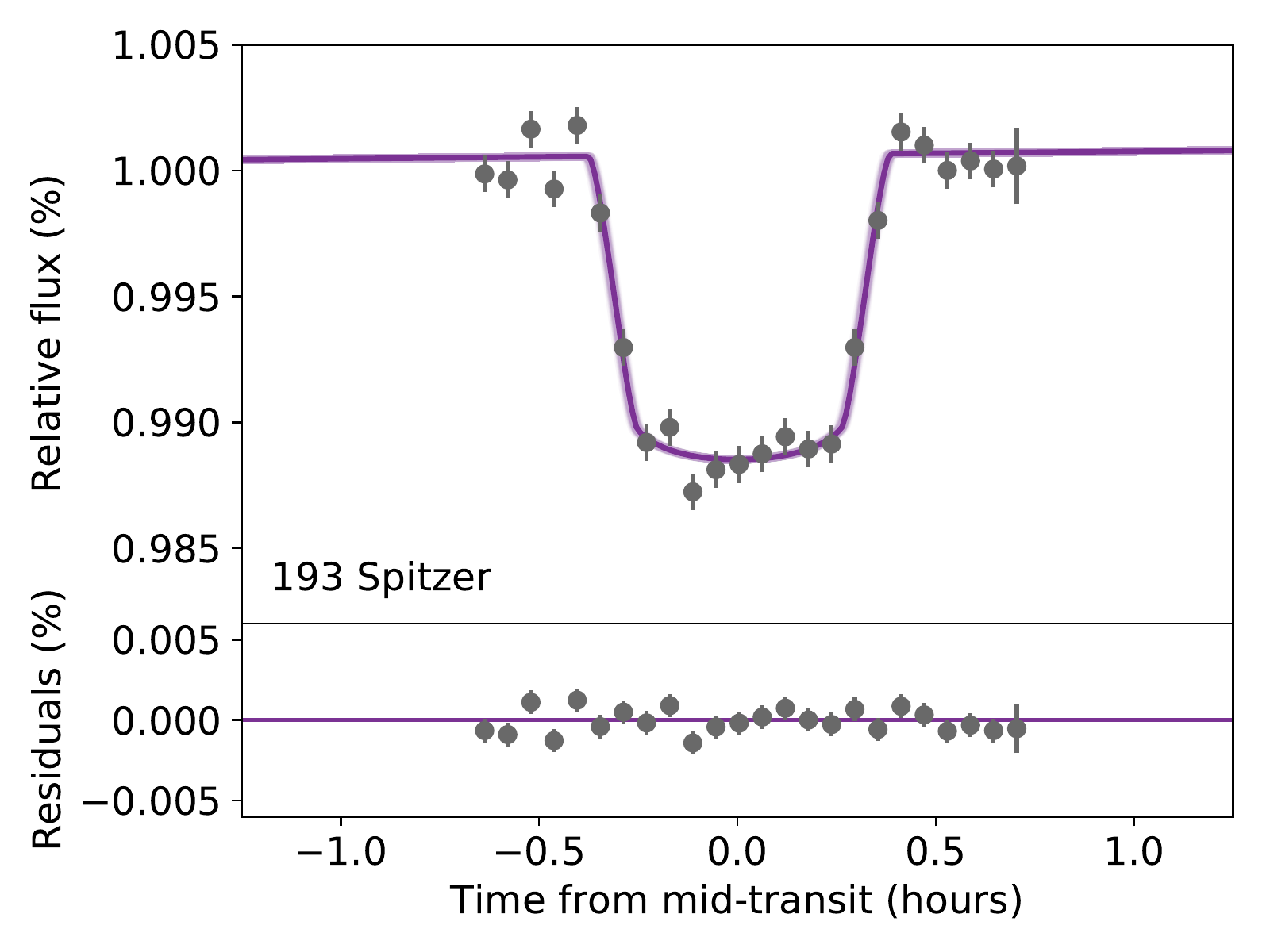}	
\includegraphics[width=0.24\textwidth]{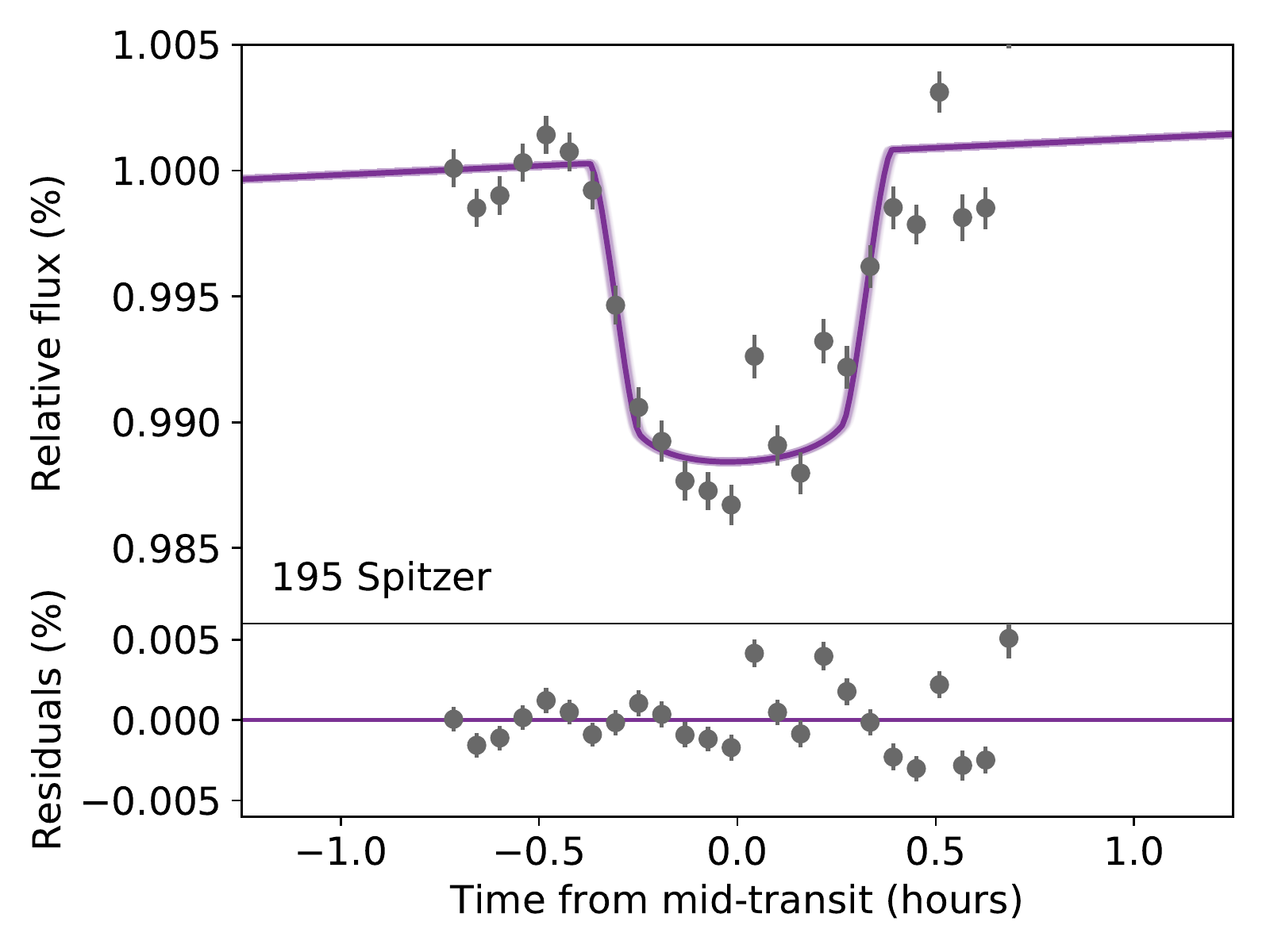}	
\includegraphics[width=0.24\textwidth]{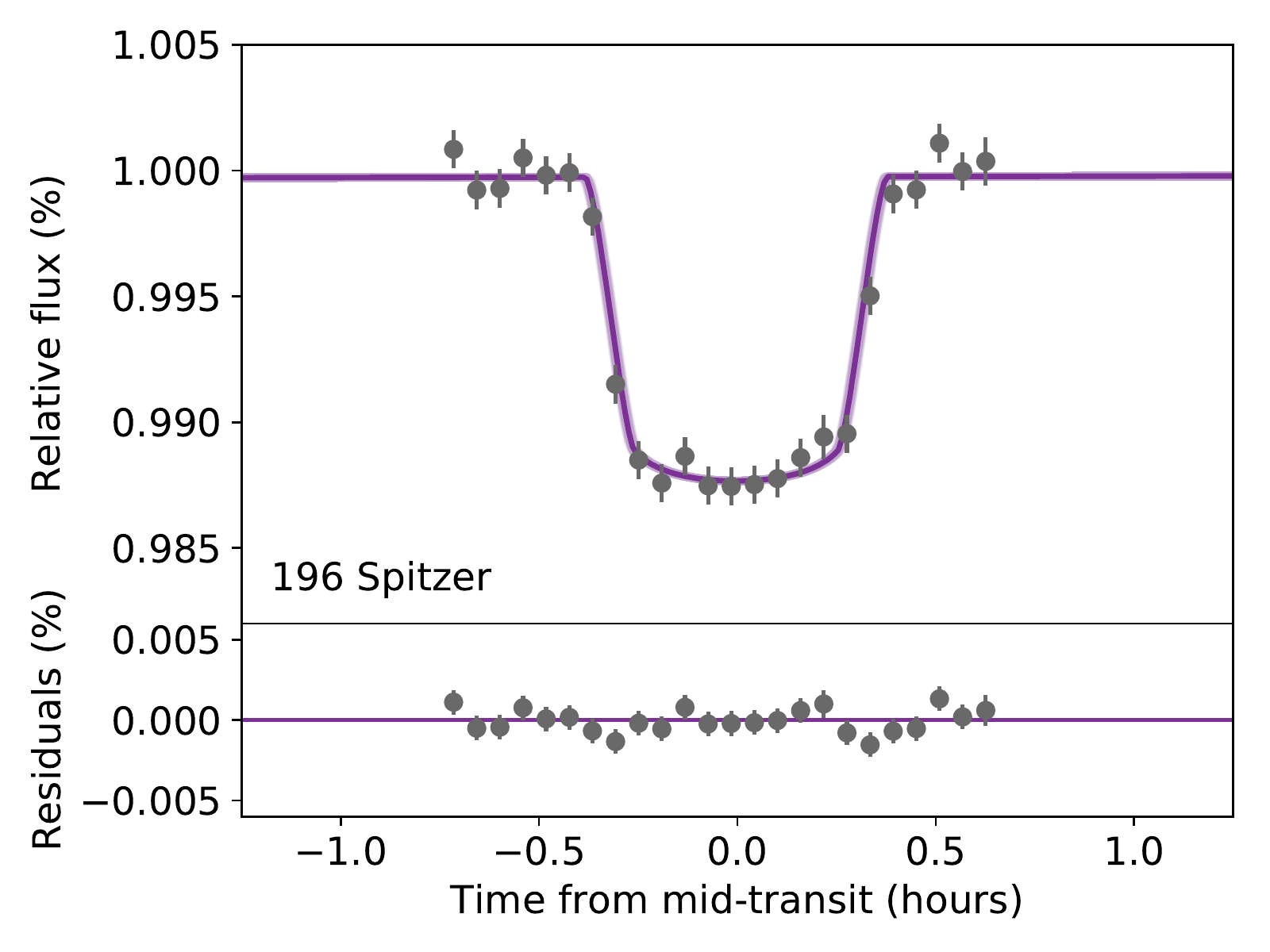}
\includegraphics[width=0.24\textwidth]{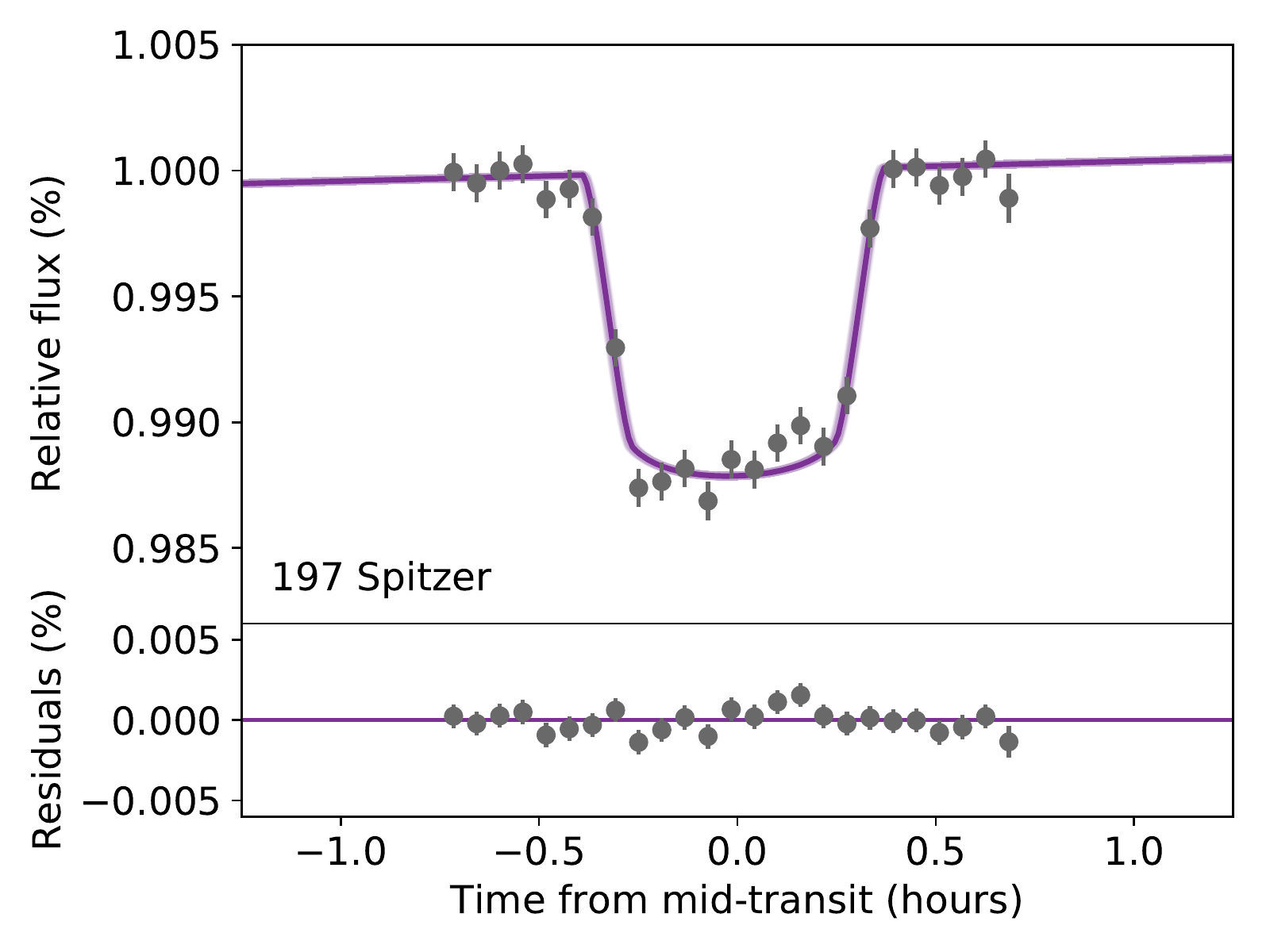}
\includegraphics[width=0.24\textwidth]{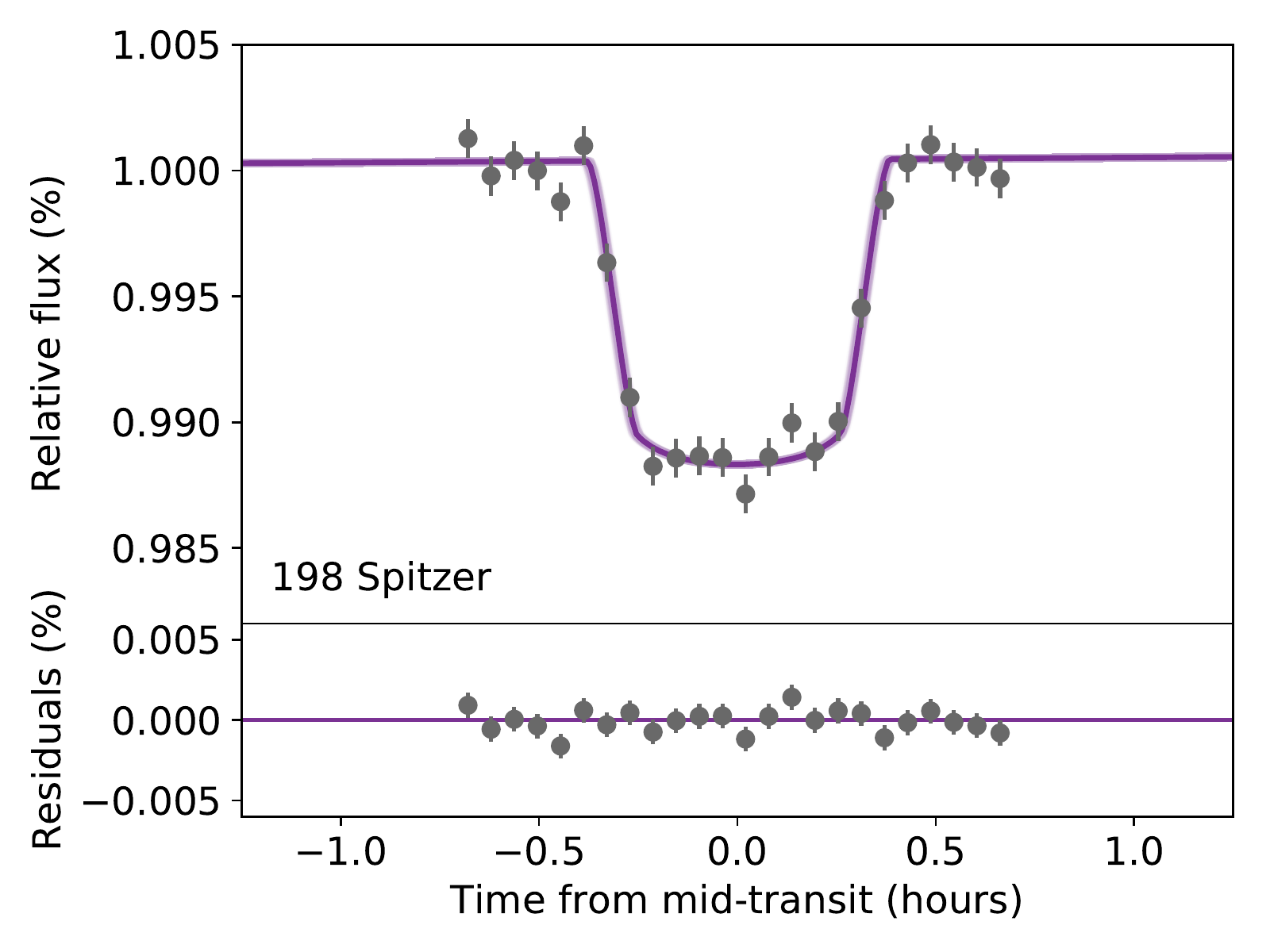}
\includegraphics[width=0.24\textwidth]{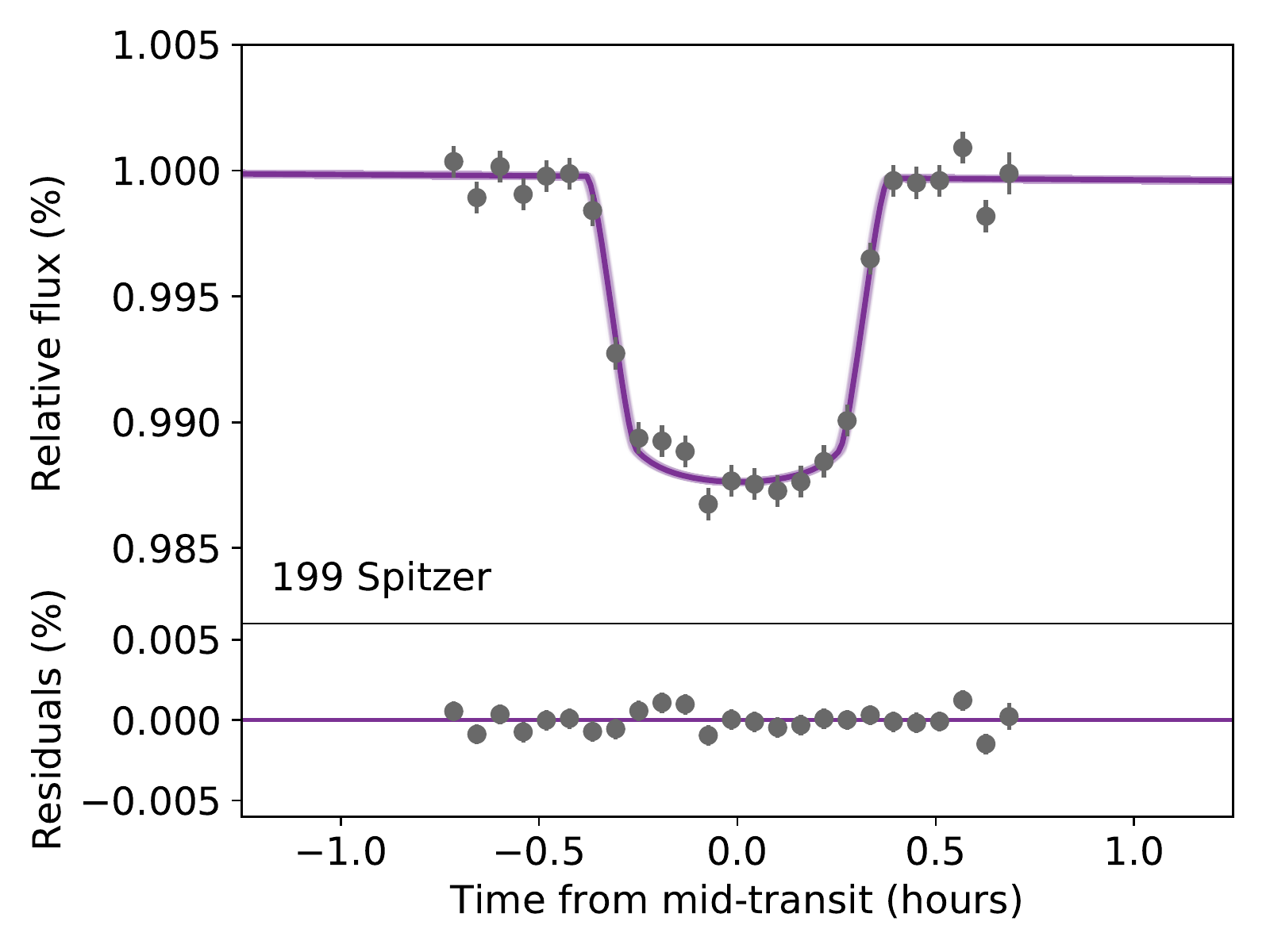}
\includegraphics[width=0.24\textwidth]{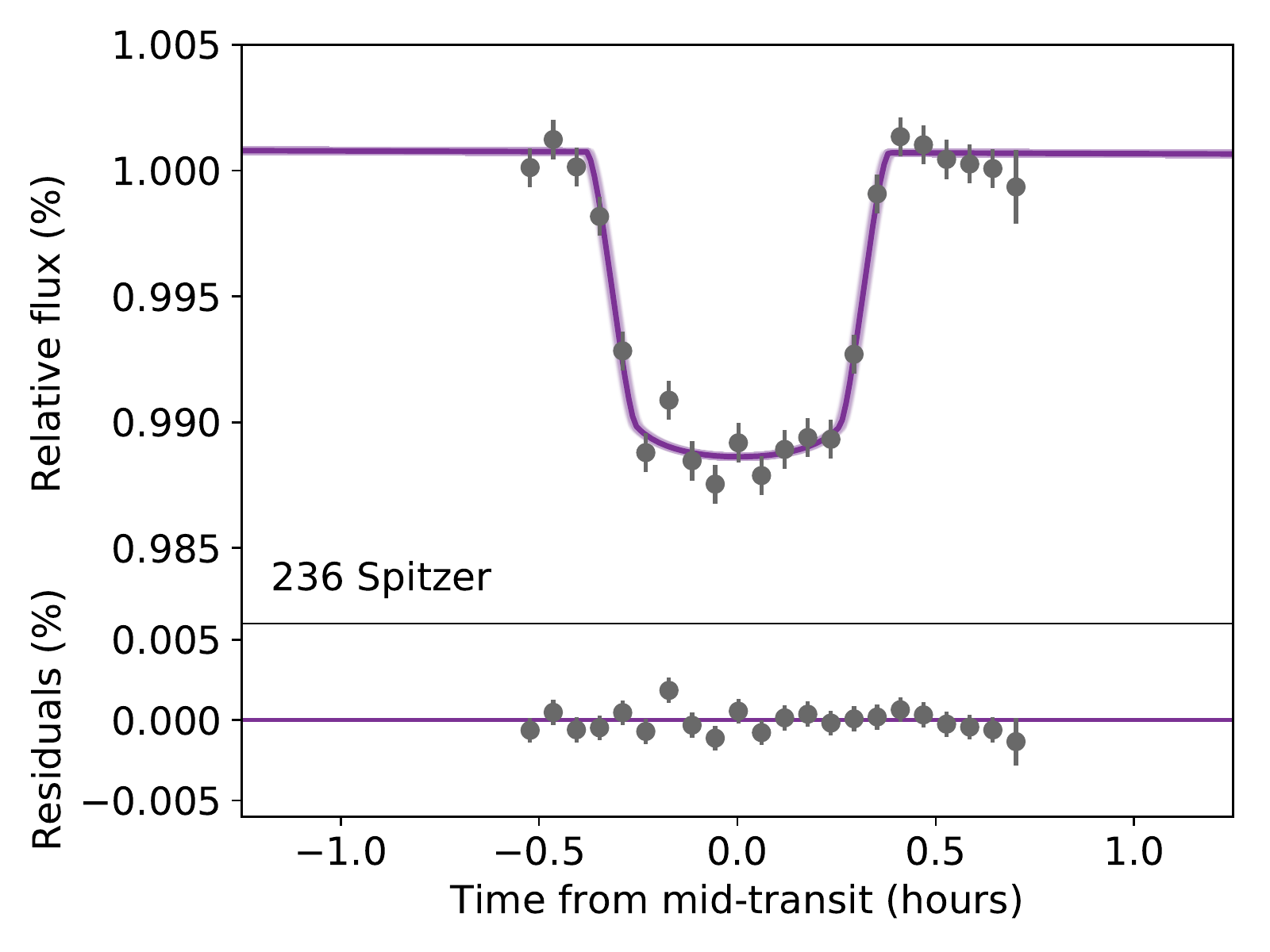}
\label{fig:all_transits}
\caption{Best fit and residuals for \textit{Spitzer} and MEarth transits from our MCMC analysis, fitting only for $T$ (Section \ref{sec:fitting}). The data are uncorrected for stellar variability. The horizontal axis is time from transit mid-point in hours and the vertical axis is the relative flux. Each light curve is centered at mid-transit time expected from a linear ephemeris and the planet period from \citet{Mann2016ZODIACALCLUSTER}. The data has been binned to 3.5 minutes. The panels are denoted by the number of transits since the ephemeris and the observatory. Note that transit 192 was observed twice, with both \textit{Spitzer} and MEarth.}
\end{figure*}

\subsection{Space-based photometry from \textit{Spitzer} and {\it K2}}

K2-25 was observed by the \textit{Kepler} spacecraft in Campaign 4, between 8 February 2015 and 20 April 2015 at 30 minute cadence. We downloaded the lightcurves available on the Mikulski Archive for Space Telescopes (MAST) on 7 September 2018. After examining the different data reductions available we use data from the K2SC algorithm \citep{Aigrain2016K2SC:Regression}. We manually inspected each transit within a four hour window to either side of the transit midpoint and discarded six transits for which there were clear signs of a flare; this inspection was completed without regards to the quality of the transit fit. 

\citet{ThaoInPrep} obtained transit observations with \textit{Spitzer}; these data and the corresponding analyses will be presented in detail in that work. To briefly summarize, {\it Spitzer} observed 10 full transits of K2-25b, five in each of 3.6$\mu$m and 4.5$\mu$m, both taken with the Infrared Array Camera \citep[IRAC, ][]{Fazio2004TheTelescope/i} over the period of 2016 November 28 to 2017 May 11 (Program ID: 13037, PI: Mann). We extracted the light curves from the {\it Spitzer} data using using a high-resolution pixel variation gain map \citep[PMAP;][]{Ingalls2012Intra-pixelIRAC} to correct for intra-pixel sensitivity variations. Fits using other techniques for high-precision photometry with {\it Spitzer}, including nearest neighbors \citep[NNBR;][]{Lewis2013ORBITALHAT-P-2b}, and pixel-level decorrelation \citep[PLD;][]{Deming2015iSPITZER/iDECORRELATION} yielded similar results, although PMAP provided the lowest red noise and most consistent transit depths at a given wavelength across the transits.


\begin{deluxetable}{ll} 
\label{tab:params} 
\tablecaption{Parameters for K2-25b used in analysis} 
\tablehead{ \colhead{Parameter} & \colhead{Value}}
\startdata
    Period (days) & $ 3.4845638 $ \\
    $T_0$ (BJD) & $ 2457062.57958 $ \\
    $r_P/R_*$ & $ 0.10807 $ \\
    $ \rho_*$ & $ 10.597 $ \\
    $ \sqrt{e}\sin{w}$ & $ 0.405 $ \\
    $ \sqrt{e}\cos{w}$ & $ 0.219 $ \\
    $ b $& $ 0.655 $ \\
    $ q_\mathrm{1,MEarth} $ & $ 0.288 $ \\
    $ q_\mathrm{2,MEarth} $ & $ 0.188 $ \\
    $ q_\mathrm{1,Spitzer CH1} $ & $ 0.124 $ \\
    $ q_\mathrm{2,Spitzer CH1} $ & $ 0.161 $ \\
    $ q_\mathrm{1,Spitzer CH2} $ & $ 0.113 $ \\
    $ q_\mathrm{2,Spitzer CH2} $ & $ 0.159 $ \\
    $ q_\mathrm{1,K2} $ & $ 0.559 $ \\
    $ q_\mathrm{2,K2} $ & $ 0.270 $
\smallskip 
\enddata 
\caption{Planetary parameters for K2-25b adopted in this work, which are the maximum of the posterior distributions from the ``combined fit'' from \citet{ThaoInPrep}. $q$ are the limb darkening parameterization from} \citet{Kipping2013EfficientLaws}.
\end{deluxetable}

\subsection{Ground-based photometry from the MEarth Observatories}

We obtained lightcurves of K2-25 with the MEarth Observatories \citep{Irwin2015TheM-dwarfs}. MEarth-North comprises eight 40 cm telescopes at Fred Lawrence Whipple Observatory on Mount Hopkins, Arizona, and MEarth-South is a near-twin located at Cerro Tololo Inter-American Observatory (CTIO) in Chile. The MEarth CCDs are 2048x2048 pixels with pixels scales of $0\farcs78$ pixel$^{-1}$ in the north and $0\farcs84$ pixel$^{-1}$ in the south. All data presented here were observed using the Schott RG715 filter, the filter profile for which is available in \citet{Dittmann2016CALIBRATIONM-DWARFS}. $60$s integration times were used throughout, and the recorded observation time is the midpoint of the exposure. Photometric monitoring began on 9 December 2015, and this work includes data obtained through 4 August 2018. We monitored the brightness of the star using a single telescope from MEarth-South, obtained a set of back-to-back exposures twice nightly, with a typical time separation of 20 minutes. We observed 12 transits using between three and eight telescopes at both the northern and southern MEarth sites, during which the star was monitored continuously with a cadence of approximately 1.5 minutes. $3$ additional transits were observed, but are not analyzed due to the photometric precision and systematics; this data quality cut was performed prior to performing any transit depth or transit time analyses.

The MEarth data are reduced with standard differential aperture photometry using a pipeline based on \citet{2007MNRAS.375.1449I} with differences described in the data release notes\footnote{https://www.cfa.harvard.edu/MEarth/DR4/processing/index.html}. The data are analyzed as described in \citet{Newton2016THENEIGHBORHOOD}; in brief, we simultaneously fit for variability induced by changes in precipitable water vapor, constant magnitude offsets, and the intrinsic stellar variability. The magnitude offsets result from reference stars being located on different parts of the detector when on either side of the meridian, a consequence of German Equatorial Mounts, and are also introduced to model otherwise unaccounted for changes in the flat-field. The stellar variability is assumed to be a sinusoid of variable amplitude and phase. We perform the simultaneously fit across a grid of rotation periods and select the best-fitting period. 

For MEarth transit observations, we independently reduce data obtained by a single telescope for a given transit. We first mask the transit, and fit the out-of-transit data to our rotation and systematics model. We fix the rotation period to the best-fitting value from our rotation analysis of data obtained from our long-term photometric monitoring, and fit only for the amplitude of the sinusoid and the scale factors that determine the contributions of the systematics. For most transits, the out-of-transit baseline is $>2$ hours both before and after the transit. We remove systematics but preserve the stellar variability. Data from all telescopes that observed a given transit are combined into a single lightcurve, preserving each datum without averaging.


\subsection{Data not analyzed}

\citet{ThaoInPrep} analyze two transits obtained from the Las Cumbres Observatory as part of their joint fit. We do not use the LCO data in the present work because the precision on the individual transits is insufficient for transit timing analyses. We also do not analyze data from {\it K2}: due to the $30$ minute sampling, neither transits nor stellar flares are fully resolved, which we found rendered modeling individual transit events challenging.

\section{Transit analysis} \label{sec:fitting}

We fit each transit independently using \texttt{batman} \citep{Kreidberg2015Batman:Python}, which implements the \citet{Mandel2002AnalyticSearches} transit model, and \texttt{emcee} \citep{Foreman-Mackey2013EmceeHammer}, an implementation of the affine-invariant Markov chain Monte Carlo (MCMC) ensemble sampler proposed by \citet{Goodman2010ENSEMBLEINVARIANCE}. 
For the MEarth data, we use data within 2.5 hours of the transit midpoint or the maximum available baseline, whichever is greater. We remove 3$\sigma$ outliers. Stellar variability has not be removed in either dataset. Our fits for the \textit{Spitzer} and MEarth transits are shown in Figure \ref{fig:all_transits}.

We use \texttt{batman} to generate our lightcurve model. The model consists of the transit midpoint ($T$); the planet's period ($P$), radius ratio ($\rprstar$), semi-major axis ratio ($a/R*$), inclination ($i$), eccentricity ($e$), argument of periastron ($\omega$); and the star's limb darkening coefficients ($q1$ and $q2$). We additionally include a constant offset ($A$) and a linear term ($B$) to account for trends in the flux levels presumed to be related to stellar rotation (the maximum observing window spans one-tenth of the stellar rotation period). All fixed parameters are set to the maximum posterior values of ``combined fit'' of \citet{ThaoInPrep}, listed in Table \ref{tab:params}. 

We perform two fits to our data. (1) We fix all parameters except for the planetary parameters $T$ and $\rprstar$, and the variables $A$ and $B$. These fits are used in Section \ref{sec:spots} to assess the potential for spots impacting our transit measurements. (2) We fix all parameters except for $T$, $A$, and $B$. These fits are used in Section \ref{sec:companions} for our transit timing analysis. 
We use a uniform prior on all parameters. We use an ensemble of 30 walkers and first run $100000$ steps with an $80000$-step  burn-in, which are discarded. The remaining $20,000$ steps are used to create our final probability distributions.

\section{Starspots on K2-25} \label{sec:spots}

\begin{figure}
\includegraphics[width=\columnwidth]{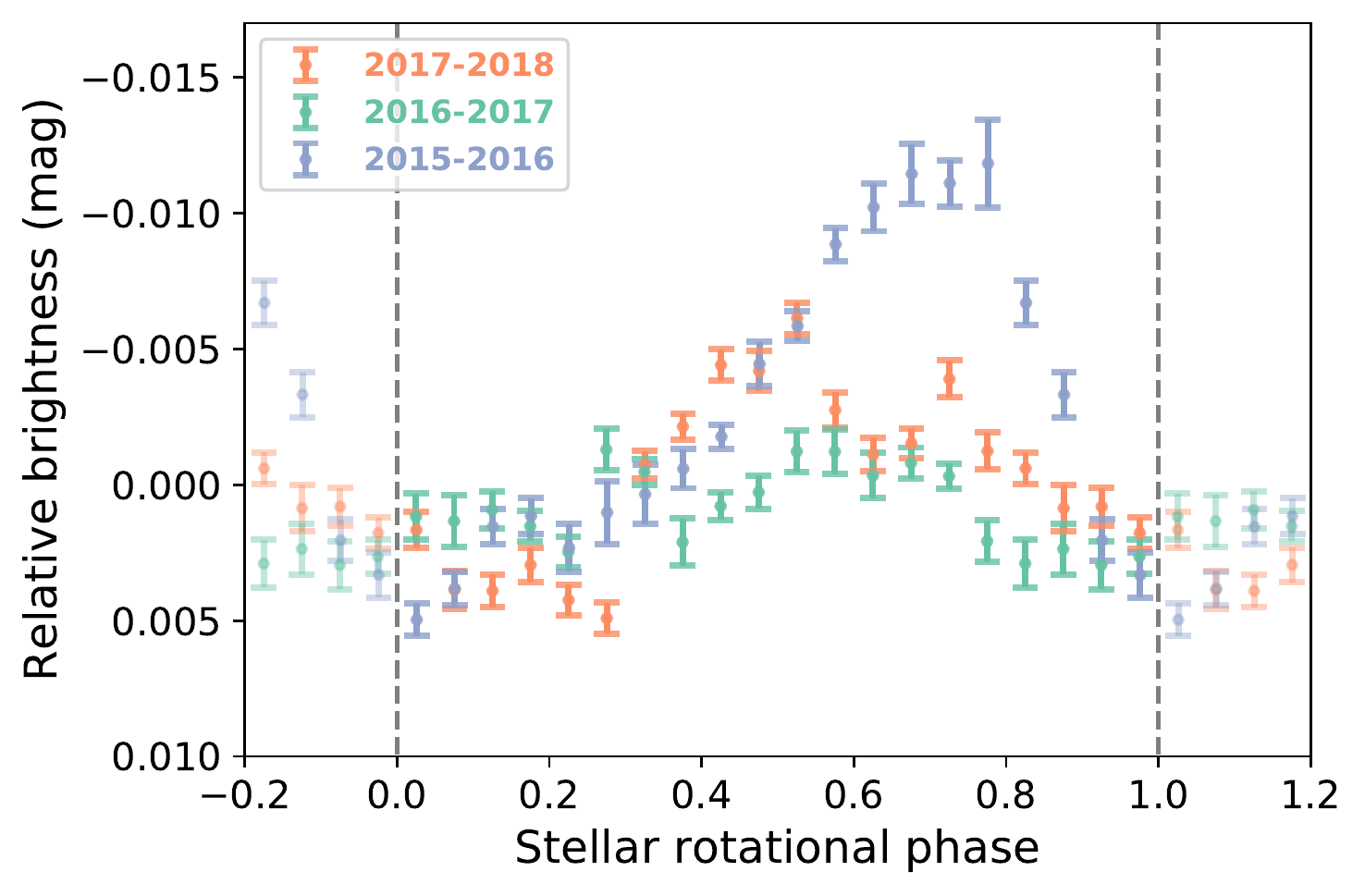}
\caption{Stellar rotational variability over three years, folded on the stellar rotation period and averaged in 20 evenly spaced bins in phase. The data are separated into three time periods that correspond to observing seasons; phase-folded data from each season are averaged separately and plotted in different colors. While our photometric monitoring for rotation spans three observing seasons from 2015-2018, the majority of our transit observations (Figure \ref{fig:depth}) were collected in the 2016-2017 observing season, which corresponds to a time of lower stellar photometric variability. \label{fig:rotation}}
\end{figure}

\begin{figure}
\includegraphics[width=\columnwidth]{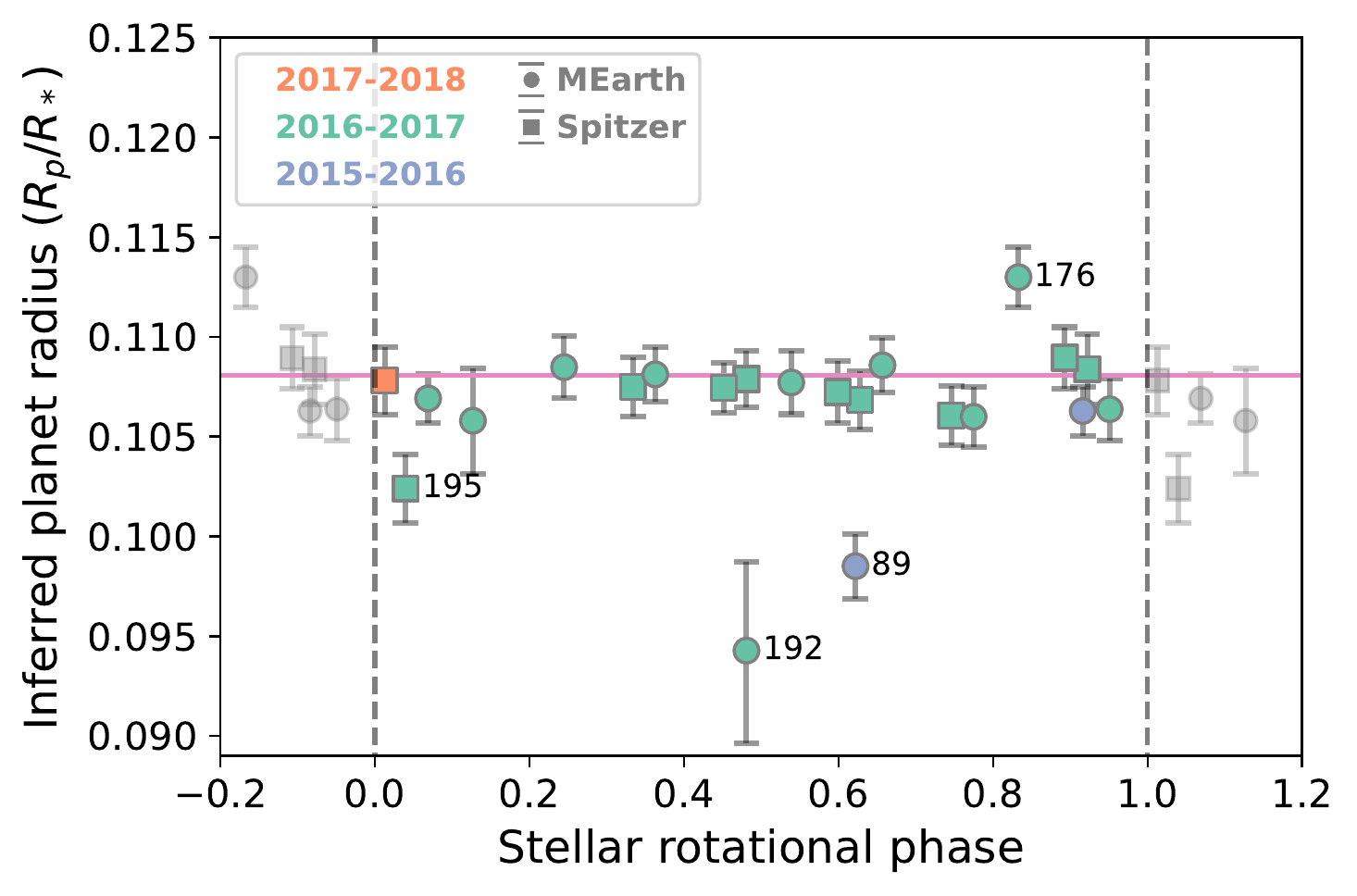}
\caption{Inferred planet radius as a function of stellar rotation phase. Each point corresponds to a different transit observation. The color of the points indicates the observing season in which the transit took place, and match the colors in Figure \ref{fig:rotation}; the majority of transits were observed in 2016-2017, which was a time of low photometric variability. Circles indicate transits obtained using MEarth, and squares are transits obtained using \textit{Spitzer}. Significant interference in the transit light curve due to spot configuration could be indicated by outliers in this diagram or by a correlation between phase and $\rprstar$. Four outliers are marked in the figure and discussed further in Section \ref{sec:spots}. No correlation is apparent. \label{fig:depth}}
\end{figure}

Stellar surface inhomogeneities, such as spots, faculae, and plages, result in variations on timescales related to the stellar rotation period, and can affect the transit light curve. For example, if the path of transit crosses a bright region, it will cause an extra drop in brightness. Individual crossings can sometimes be clearly discerned in a lightcurve \citep{Dittmann2009ATrES-1, Sanchis-Ojeda2011StarspotsSystem}. However, if crossings of dark (or bright) regions are not identified and attributed to the transiting planet, one will conclude that the planet has a smaller (or larger) radius than it actually does. In the models in \citet[Table 5]{Rackham2017ThePlanets}, a hot Neptune transiting an M4V star with solar-like spots and faculae (the scenario producing the largest stellar signal) appears to have $\rprstar$ larger than the unspotted case by  $\rprstar=0.005$.

\subsection{Long-term variability}

Figure \ref{fig:rotation} shows the long-term variability of K2-25, with phase-folded lightcurves from MEarth. The star has a rotation period of $1.88$ days and a peak-to-peak photometric amplitude of close to 2\% in the MEarth lightcurve (it is similar in \textit{K2}). This is typical for rapidly rotating mid M dwarfs \citep{Douglas2014THEHYADES, Newton2016THENEIGHBORHOOD}. The photometric amplitude remained high throughout the $90$ day \textit{K2} campaign and the first year (2015-2016) of ground-based monitoring from MEarth, and maintained a consistent morphology.  

The variability models from \citet{Rackham2017ThePlanets} suggest that the 2\% variability could be explained by either a $\gtrsim 30\%$ covering fraction of small spots, or a $\lesssim10\%$ covering fraction of large spots. In 2016-2017, the phase shifts and the amplitude decreases by half, and in 2017-2018 the amplitude increases slightly and acquires a double dip. This could result from fewer spots, but as noted in \citet{2018ApJ...865..142B}, the decrease in photometric variability could instead correspond to a rearrangement in spots.


\subsection{Transit depth variations and spot crossings}

In Figure \ref{fig:depth}, we consider the inferred planet radius $\rprstar$ as a function of stellar rotational phase. The results in this figure derive from our fit varying both $\rprstar$ and $T$. 
Most transit data from MEarth and all data from \textit{Spitzer} were obtained when K2-25 was in a low-variability phase. The data show consistent transit depths across most transits: if our transit depths are impacted by stellar surface inhomogeneities, there is not a strong time-dependent component. Four outliers are denoted by the transit number in Figure \ref{fig:depth} and discussed further in this section. We also generally found in-transit deviations to be similar to the out-of-transit variations in all transits except for MEarth 87, which we comment on below.  

For two outliers, we suggest the transit depth differences are due to unaccounted for errors. MEarth 176, for which the transit is deep, has strong systematic variability visible in the post-transit data, which could be impacting the depth measurement. 
MEarth 192, for which the transit is shallow, is our noisiest lightcurve and a simultaneous \textit{Spitzer} transit has a normal transit depth. 

\begin{figure}
    \centering
    \includegraphics[width=0.95\columnwidth]{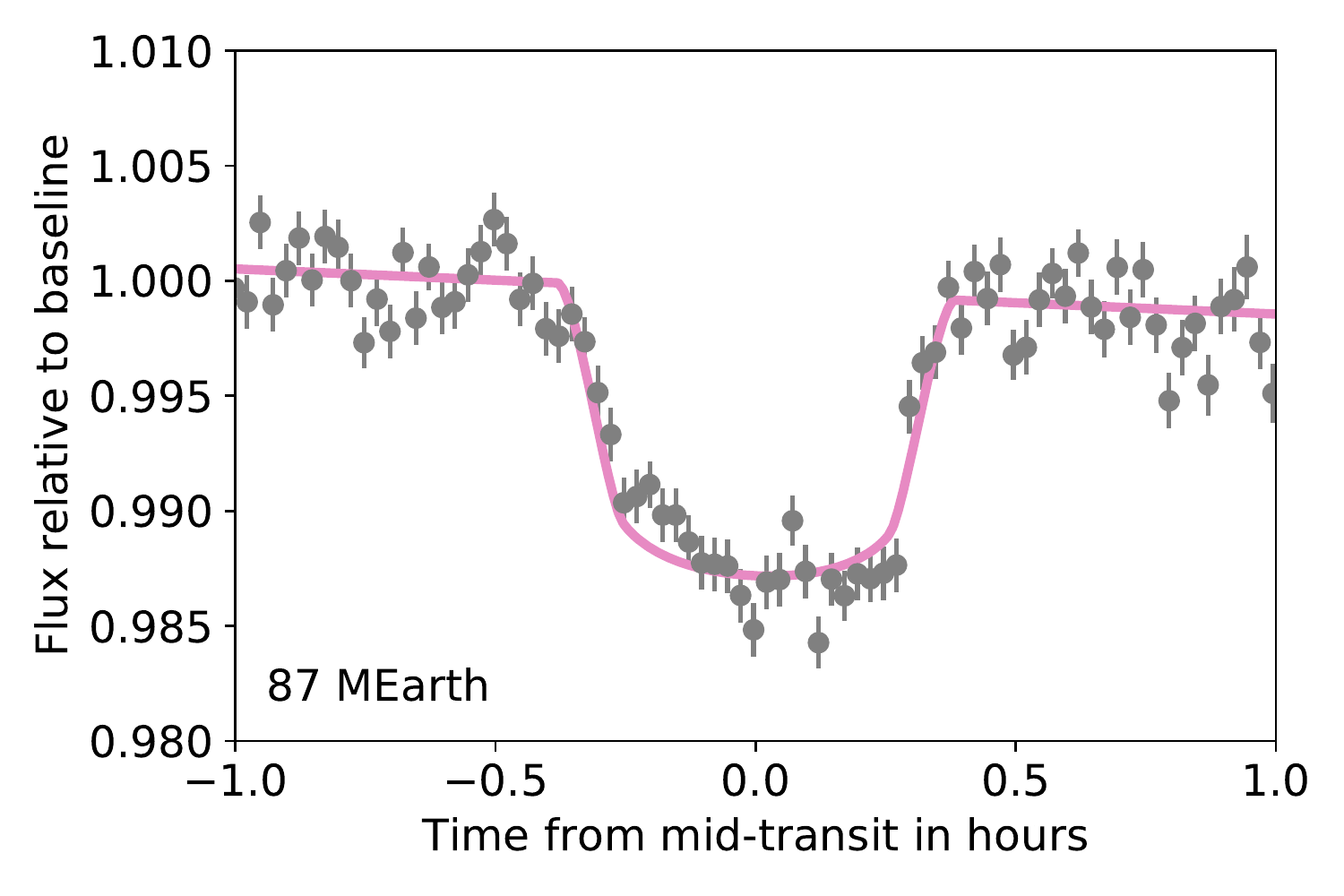}
    \includegraphics[width=0.95\columnwidth]{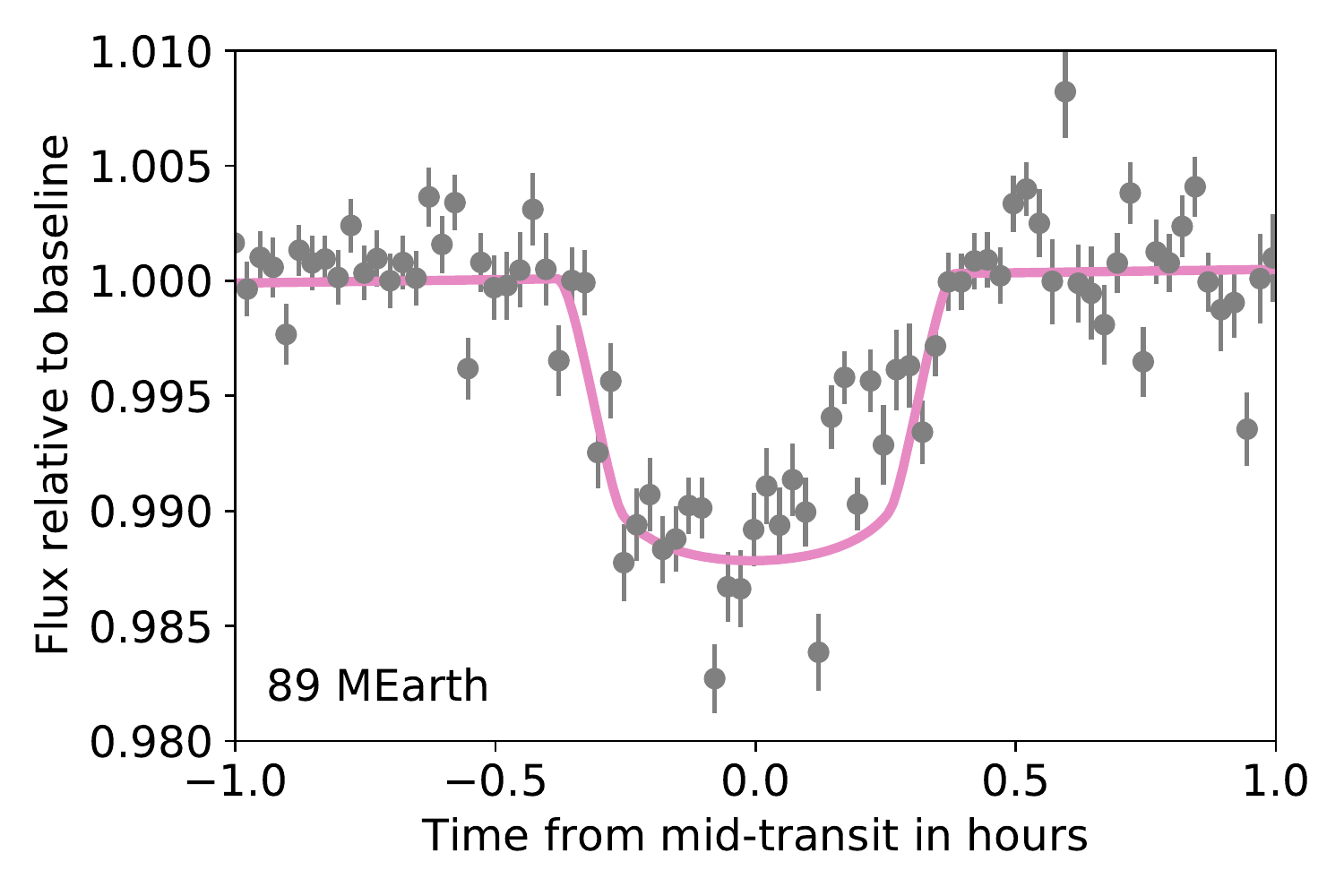}
    \includegraphics[width=0.95\columnwidth]{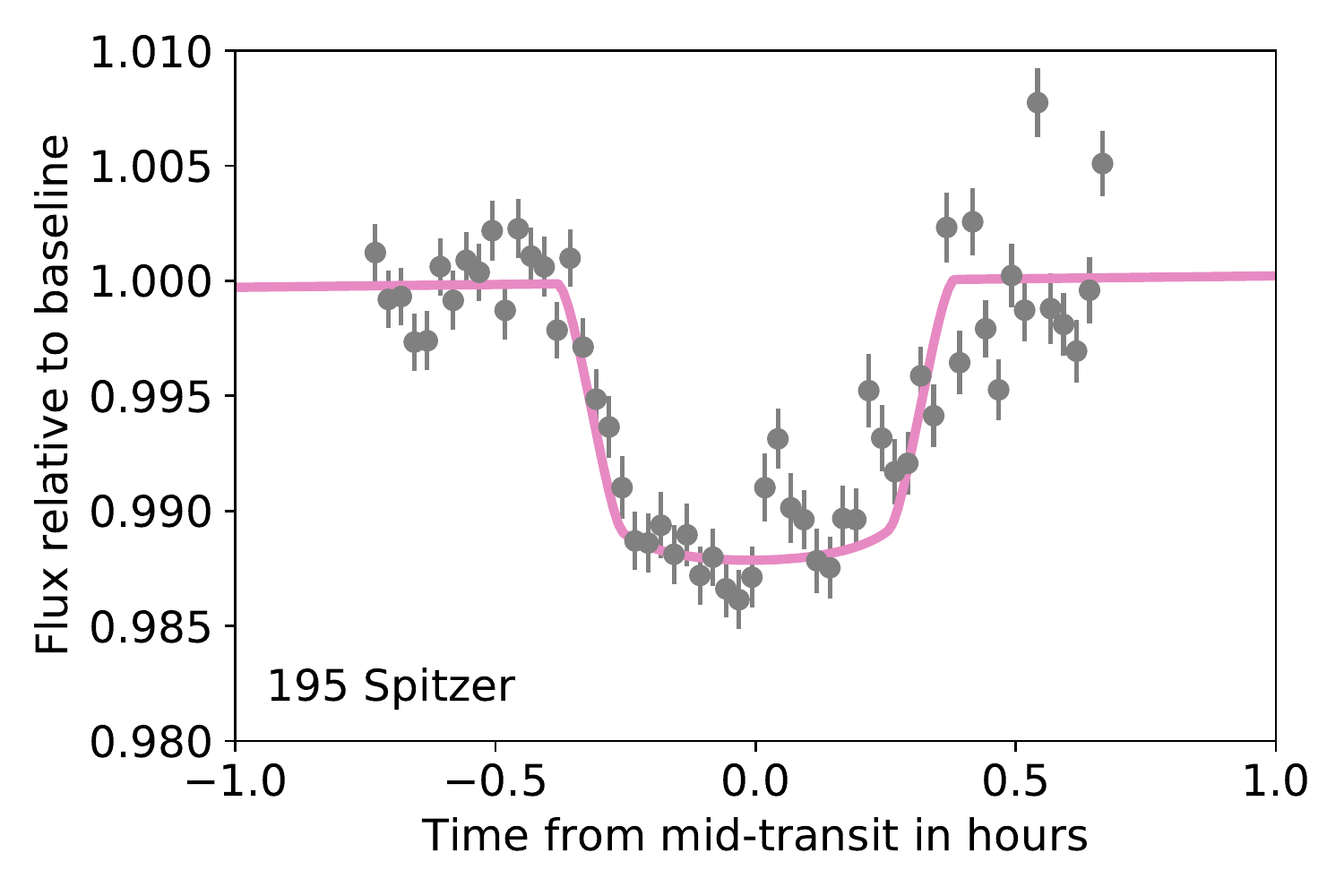}
    \caption{Transit lightcurves of the three transits that show deviations from the transit model. MEarth 87 is shown on the top, MEarth 89 in the middle, and {\it Spitzer} 195 on the bottom. The data are binned using a robust weighted mean to $1.5$ minutes. The model shown is the maximum posterior} of the combined fit model from \citet{ThaoInPrep}, adjusted by our fits for $A$ and $B$ to account for overall changes in slope.
    \label{Fig:spot_transits}
\end{figure}

For the other two outliers, spots may be responsible for the transit depth differences. These transits are shown in Figure \ref{Fig:spot_transits}, and compared to maximum posterior fit from the ``combined'' transit model from \citet{ThaoInPrep}. MEarth 89 corresponds to one of only two transits observed 2015-2016, during K2-25's high-variability phase. The transit is noticeably shallower than the model from \citet{ThaoInPrep}. This could indicate that the transit chord at this moment is darker than the unocculted stellar surface. However, we note that the post-transit data is impacted by scatter beyond the photometric errors. Also shown in Figure \ref{Fig:spot_transits} is the other transit obtained in 2015-2016 (MEarth 87), which has a transit depth consistent with the others, but shows an asymmetry during ingress. Finally, in {\it Spitzer} 195, we see a candidate crossing of a dark spot. However, during this transit, the target did not fall on the ``sweet spot’’ of the {\it Spitzer} detector; and uncharacterized systematics are of concern. 

While the measured transit depths are not significantly different, we also noted candidate spot crossings in MEarth 180. However, systematics are again of concern.

\begin{figure*}
\plotone{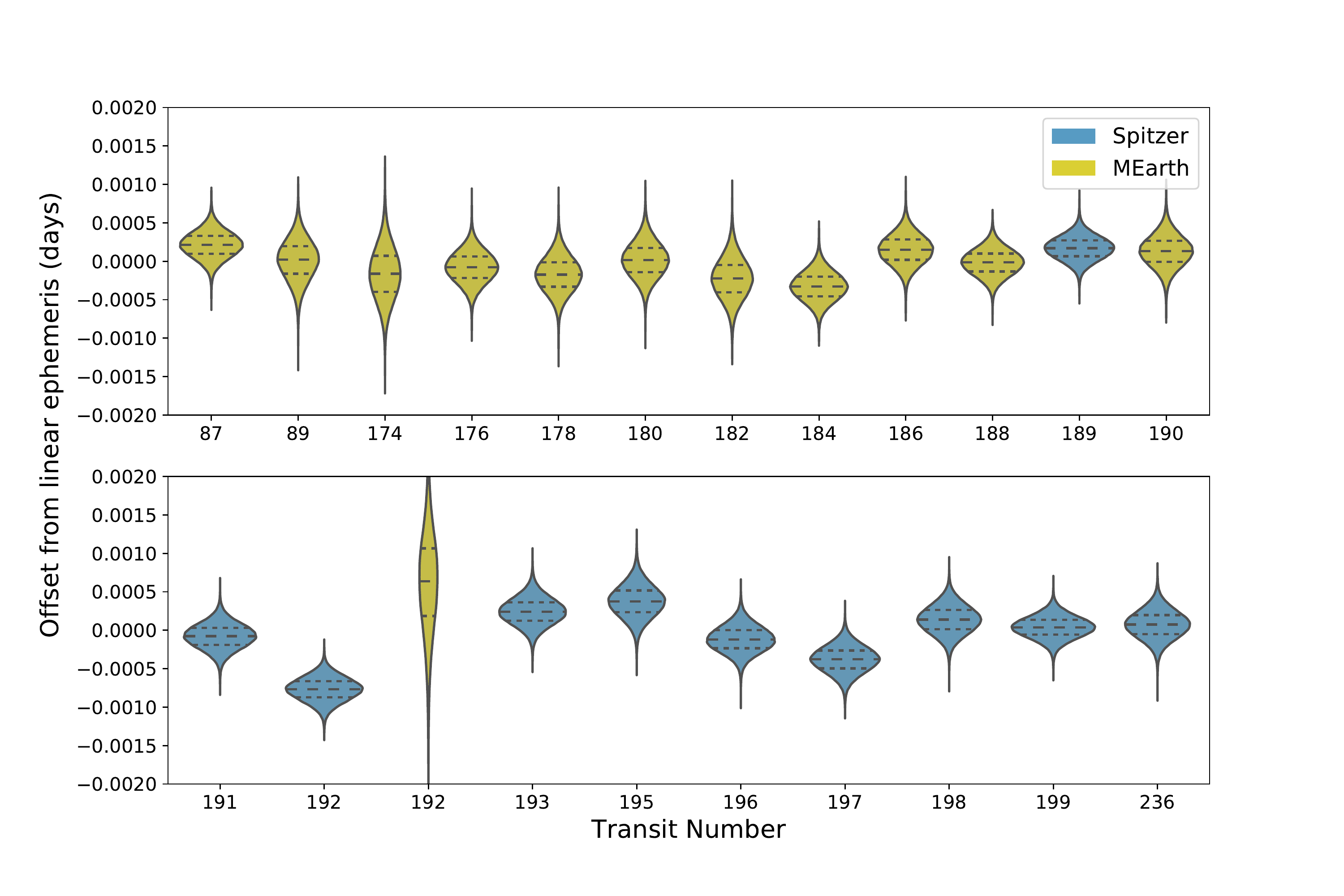}
\label{fig:violin}
\caption{Violin plot showing distribution of simulated mid-transit times (in days) from MCMC analysis. Transits are shown individually (denoted by the transit number, counting up from the transit ephemeris given in \citealt{Mann2016ZODIACALCLUSTER}), and the width of the distribution indicates the probability density of the mid-transit time as that vertical location. The dashed line marks the $50^{th}$ percentile, and the dotted lines show the $16^{th}$ and $84^{th}$
percentiles. The spread of the distribution for MEarth transit 192 can be attributed to moderate data quality.}
\end{figure*}

\section{Constraints on companions} \label{sec:companions}


An unperturbed planet orbiting its star will take the same amount of time to complete each revolution around its star, with small exceptions for orbital precession or tidal decay. However, when a companion is added to the system, mutual gravitational interaction causes the two (or more) planets to exchange energy and angular momentum. This causes short-term oscillations of the semimajor axes and eccentricities of these planets, which in turn leads to variations in the time interval between transits. 
For example, the first positive detections of transit timing variations in an exoplanetary system was observed for Kepler-19b \citep{Ballard2011TheVariations}, which  consists of a 2.2 R$_\oplus$  planet on a 9.3-day orbit around a Sun-like star. The transit times of Kepler-19b showed sinusoidal like variations indicating the presence of at least one non-transiting companion at a longer period. 


\subsection{Planet mass estimation}

An approximate planet mass is necessary for TTV analysis. We use the nonparametric mass--radius relationship from \citet{2018ApJ...869....5N} implemented in the publicly available package \texttt{MRExo}\footnote{\url{https://github.com/shbhuk/mrexo}} \citep{2019arXiv190300042K}. We use the provided results from {\it Kepler} dataset to predict the planet's mass given its radius; this dataset includes exoplanets with mass from both radial velocities and N-body dynamical fits to TTV. Using the planetary radius and error from \citet[]{ThaoInPrep}, we estimate a mass of $7^{+10}_{-4}$ $M_\oplus$ for K2-25b, where the error bars represent the 68\% confidence interval. The mass-radius relation for sub-Neptunes from \citet{Wolfgang2016PROBABILISTICPLANETS}, $M/M_\oplus = 2.7(R/R_\oplus)^1.3 )$, yields $M_P=13.5\pm1.9$ $M_\oplus$ where the error is dominated by the intrinsic scatter in the relationship. We note that when looking at planets in the sample from \citet{2018ApJ...869....5N} between $4$ and $5$ $R_\oplus$, there are a few outliers to the distribution at higher masses; while we adopt the mass from the \citet{2018ApJ...869....5N} mass--radius relation, we also comment on our expectations if K2-25b were to have a mass of $30$ $M_\oplus$.


\subsection{Transit timing analysis} \label{subsec:ttvs}

We investigated whether the transit times of K2-25b, which we display in Figure \ref{fig:violin} for the {\it Spitzer} and MEarth data, are consistent with an unperturbed linear ephemeris or whether there is evidence for additional near-resonant companions in this system. We follow a procedure similar to \citet{Dittmann2017ACampaign}.

Our observations comprise 39 individual transit observations over 822 days (236 orbits of K2-25b). The combination of ground based MEarth transit measurements and space based \textit{Spitzer} measurements allow us to measure transit times of closely-spaced transits at high precision. This dense sampling, combined with a long baseline from \textit{K2} to establish the average period of K2-25b allows us to fully sample the transit timing signal and assess whether there is evidence for another perturbing body.

The largest variation from a linear ephemeris present in our data is approximately 2 minutes in magnitude, and we see no evidence for a significant variation from a linear ephemeris. In order to determine what mass bodies we may exclude via transit timing, we use the \texttt{TTVFaster} code \citep{Deck2016TRANSITRESONANCES}, an extension of \texttt{TTVFast} \citep{Deck2014TTVFast:PROBLEMS} capable of estimating a TTV signal with eccentric planets. 

We limit our TTV sensitivity analysis to be between periods of 0.5 days and 25 days (just beyond 7 times the period of K2-25b). For each trial period, we initialize 500 versions of K2-25b drawn from the best fit parameters determined by \citet{ThaoInPrep} we have presented here. For each of these 500 systems, we initialize 500 versions of a possible K2-25c. Each companion is drawn to have an eccentricity between 0 and 0.5 with uniform probability, an inclination angle relative to the star between 80 and 110 degrees with uniform probability, and random longitudinal nodes, argument of periastron, and mean anomalies. Intially, all perturbing planet masses are 0.5 M$_\oplus$. 

For each system generated at each period, we sample the transit times of K2-25b at the timestamps of the observed data. We then fit a linear ephemeris to the generated transit times.
This effectively marginalizes over planetary orbital parameters. If the 90th percentile of the amplitude of these transit timing variations is greater than the 2 minute deviation we observe in our data, then we adopt this mass value as our limiting value. Otherwise, we increase the mass of the perturber by 0.5 M$_\oplus$ and repeat the procedure until a sufficiently large transit timing variation is observed.

\begin{figure}
\plotone{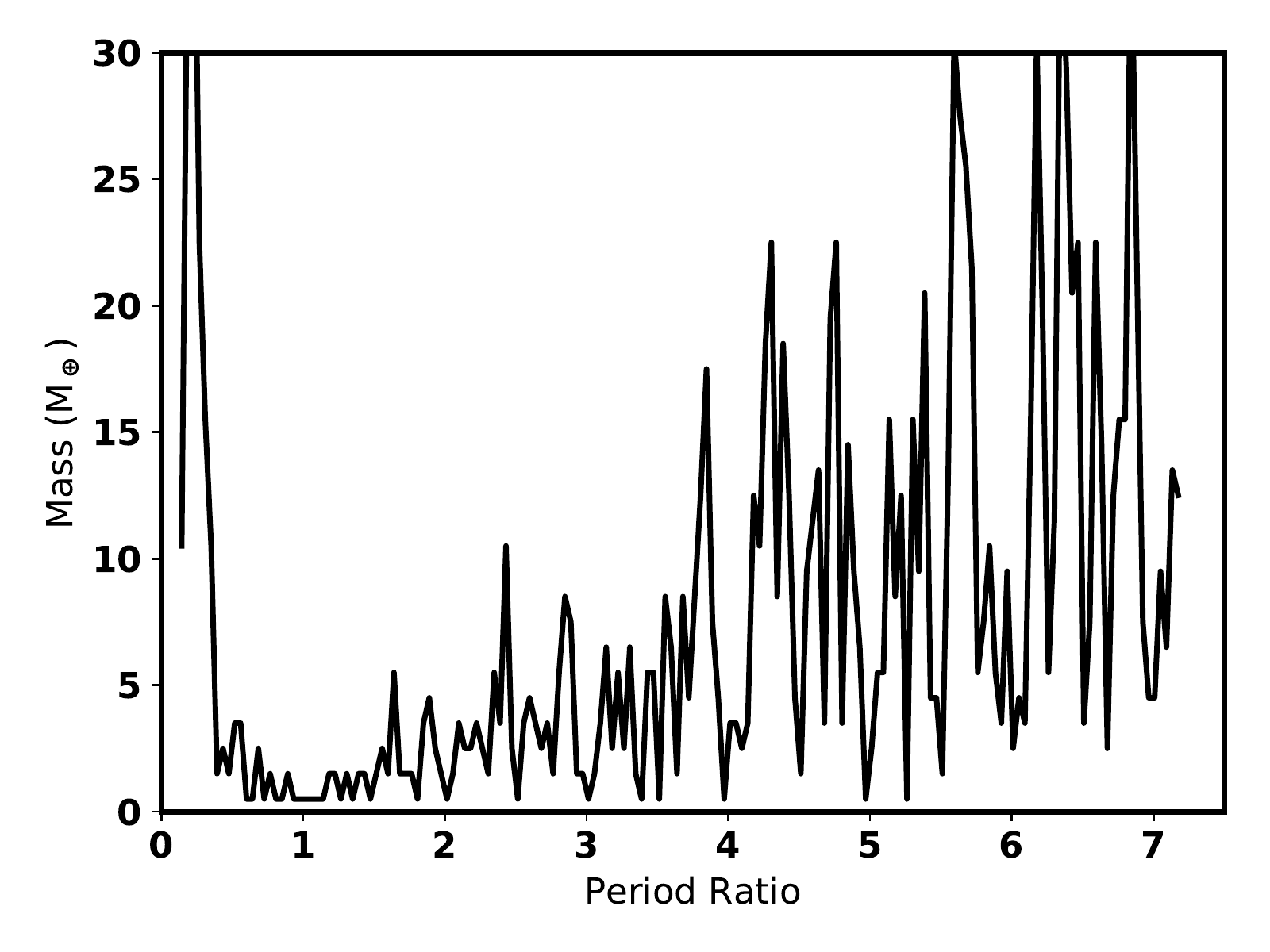}
\label{fig:TTV}
\caption{The mass of a perturbing object versus orbital period ratio required to create a 2 minute transit timing variation signal in K2-25b. We marginalize over eccentricity, orbital inclination, longitudinal node, argument of periapsis, and mean anomaly of the orbit using the \texttt{TTVFaster} code. We find that companions out to the 2:1 mean motion resonance are inconsistent with the observed transit times from MEarth, \textit{Spitzer}, and K2. Super-Earth mass companions and smaller can be consistent with our transit timing data out to $\sim$ 7:2 period ratios. Beyond the 7:2 resonance, only companions in integer period ratios with K2-25b can be ruled out.}
\end{figure}

We plot these maximum allowed perturbing mass as a function of the period ratio between a potential perturber and K2-25b in Figure \ref{fig:TTV}. We find that the lack of transit timing variations in K2-25b precludes all but the shortest period companions interior to K2-25b. Furthermore, we can largely eliminate low-mass companions out to the 2:1 orbital resonance with K2-25b. 
Beyond orbital period ratios of 7:2 we require much higher mass perturbers to create a detectable TTV signal, and are largely insensitive to possible small bodies outside of strong integer resonances. We performed the same calculation using transit times measured from fits to individual {\it K2} transits and arrived at consistent results: the {\it K2} transit times are not precise enough to strongly impact our sensitivity to TTVs.

\begin{deluxetable}{ccc}
\tablecaption{Mid-transit times}
\label{tab:table2}
\tablehead{
        \colhead{Transit Number} &
        \colhead{Mid-Transit Time (BJD)} &
        \colhead{Telescope}
        }
\startdata
$ 0 $ & $ 2457062.581263^{ +0.000348}_{ -0.000337} $ & K2 \\  
$ 1 $ & $ 2457066.065295^{ +0.000342}_{ -0.000300} $ & K2 \\  
$ 2 $ & $ 2457069.547610^{ +0.000373}_{ -0.000366} $ & K2 \\  
$ 3 $ & $ 2457073.034524^{ +0.000392}_{ -0.000385} $ & K2 \\  
$ 4 $ & $ 2457076.517230^{ +0.000328}_{ -0.000329} $ & K2 \\  
$ 5 $ & $ 2457080.003011^{ +0.000368}_{ -0.000346} $ & K2 \\  
$ 6 $ & $ 2457083.486753^{ +0.000389}_{ -0.000364} $ & K2 \\
$ 10 $ & $ 2457097.424917^{ +0.000371}_{ -0.000373} $ & K2 \\  
$ 12 $ & $ 2457104.394692^{ +0.000387}_{ -0.000372} $ & K2 \\ 
$ 14 $ & $ 2457111.364489^{ +0.000385}_{ -0.000388} $ & K2 \\  
$ 15 $ & $ 2457114.847241^{ +0.000447}_{ -0.000386} $ & K2 \\  
$ 16 $ & $ 2457118.333371^{ +0.000348}_{ -0.000346} $ & K2 \\  
$ 17 $ & $ 2457121.818849^{ +0.000557}_{ -0.000598} $ & K2 \\  
$ 18 $ & $ 2457125.300855^{ +0.000290}_{ -0.000293} $ & K2 \\  
$ 20 $ & $ 2457132.270650^{ +0.000475}_{ -0.000417} $ & K2 \\  
$ 87 $ & $ 2457365.736862^{ +0.000172}_{ -0.000170} $ & MEarth \\  
$ 89 $ & $ 2457372.705796^{ +0.000273}_{ -0.000255} $ & MEarth \\  
$ 174 $ & $ 2457668.893487^{ +0.000344}_{ -0.000345} $ & MEarth \\  
$ 176 $ & $ 2457675.862698^{ +0.000204}_{ -0.000202} $ & MEarth \\  
$ 178 $ & $ 2457682.831729^{ +0.000230}_{ -0.000233} $ & MEarth \\  
$ 180 $ & $ 2457689.801044^{ +0.000227}_{ -0.000231} $ & MEarth \\  
$ 182 $ & $ 2457696.769930^{ +0.000260}_{ -0.000265} $ & MEarth \\  
$ 184 $ & $ 2457703.738953^{ +0.000186}_{ -0.000188} $ & MEarth \\  
$ 186 $ & $ 2457710.708559^{ +0.000195}_{ -0.000194} $ & MEarth \\  
$ 188 $ & $ 2457717.677520^{ +0.000172}_{ -0.000169} $ & MEarth \\  
$ 189 $ & $ 2457721.162269^{ +0.000155}_{ -0.000152} $ & Spitzer \\  
$ 190 $ & $ 2457724.646792^{ +0.000202}_{ -0.000202} $ & MEarth \\  
$ 191 $ & $ 2457728.131147^{ +0.000164}_{ -0.000161} $ & Spitzer \\  
$ 192 $ & $ 2457731.615021^{ +0.000158}_{ -0.000153} $ & Spitzer \\  
$ 192 $ & $ 2457731.616424^{ +0.000679}_{ -0.000628} $ & MEarth \\  
$ 193 $ & $ 2457735.100594^{ +0.000173}_{ -0.000178} $ & Spitzer \\  
$ 195 $ & $ 2457742.069854^{ +0.000210}_{ -0.000210} $ & Spitzer \\  
$ 196 $ & $ 2457745.553921^{ +0.000168}_{ -0.000176} $ & Spitzer \\  
$ 197 $ & $ 2457749.038225^{ +0.000174}_{ -0.000170} $ & Spitzer \\  
$ 198 $ & $ 2457752.523305^{ +0.000185}_{ -0.000186} $ & Spitzer \\  
$ 199 $ & $ 2457756.007769^{ +0.000142}_{ -0.000142} $ & Spitzer \\  
$ 236 $ & $ 2457884.936644^{ +0.000181}_{ -0.000182} $ & Spitzer \\  
\smallskip 
\enddata 
\end{deluxetable}  




\section{Discussion} \label{sec:discussion}


The hot Neptune K2-25b arrived at and appears to have thus far maintained a close-in, moderately eccentric orbit. Eccentric systems may quickly circularize due to tidal forces. If the circularization timescale for this planet is close to or larger than the age of the system, then we are likely observing this planet in the act of circularizing after some single dynamic event early in its lifetime. If, however, this timescale is much less than the age of the system, then it could be that the eccentricity is being actively excited.

\subsection{Circularization timescale} \label{subsec:circ}

We use the mass estimation determined in the previous section to calculate the circularization timescale of this system, the approximate amount of time it would take tidal forces to damp K2-25b's eccentricity. We use the equation derived in \citet{Goldreich1966QSystem} as presented in Equation 2 of  \citet{Jackson2009OBSERVATIONALEXOPLANETS}, with a negative sign on the second term due to the star's rapid spin:

\begin{equation}
\tau_\mathrm{circ} = a^{13/2}\left(\frac{63}{4}\sqrt[]{GM_*^3}\frac{R_P^5}{Q_P^\prime M_P}  - \frac{225}{16}\sqrt[]{G/M_*}\frac{R_*^5 M_P}{Q_*^\prime}\right)^{-1}\label{eq:tcirc}
\end{equation}

where $M_P$ is planet mass, $R_P$ is planet radius, $M_*$ is stellar mass, and $a$ is semimajor axis. Values of $Q_P^\prime$ and $Q_*^\prime$, the modified tidal quality factors of the planet and the star ($Q^\prime=2Q/3k$), are not well-constrained.  $Q_*^\prime$ values inferred from the circularization of hot Jupiters and stellar binaries vary \citep[see discussion and references in][]{2018haex.bookE..24M}. 
For K2-25, the tides raised by the star on the planet dominate and the term involving $Q_*^\prime$ is negligible. 
 

 We estimate a circularization timescale of $410$ Myr using $Q_P^\prime = 5\times10^4$ and the mass as determined in the previous section. Based on the 68\% confidence interval of the masses, the 68\% confidence interval of the circularization timescale is $180$ Myr to $1$ Gyr. Our selected $Q_P^\prime$ is based on Neptune's $Q_N/k$ value of $2.2\times10^4 < Q_N < 9\times10^4$ from \citet{Zhang2008OrbitalSystem}. Varying $Q_P^\prime$ by an order of magnitude adjusts the calculated timescale by an order of magnitude as well. 
 If K2-25b is in fact a higher density world, we would expect its mass to be greater and its $Q_P^\prime$ to be lower. For $M_P=30$ $M_\oplus$ and $Q_P^\prime=100$, the circularization timescale is only $4$ Myr.

We note that the K2-25 system is unlike many other planetary systems in that the star is rotating faster ($P_*=1.88$ days) than the planet orbits ($P_P=3.48$ days), driving the negative sign on the second term in Equation \ref{eq:tcirc}. An extended period of rapid rotation like K2-25's is common to all mid-to-late M dwarfs, though most known M dwarf planet hosts are presently slowly rotating \citep{Newton2016THENEIGHBORHOOD}.  $Q_*$ is a parameterization of the tidal response of the star, and is influenced by the frequencies at which the star spins and the planet orbits \citep{Ogilvie2007TidalStars}. In some extreme cases, the planetary eccentricity can be excited rather than damped \citep{Dobbs-Dixon2004Spin-OrbitPlanets}. Using the equations in \citet{Dobbs-Dixon2004Spin-OrbitPlanets}, we find that K2-25 is not in this regime; a several Jupiter mass planet would be required. 


\subsection{Migration mechanisms}
\label{subsec:migration}

The eccentric orbit of K2-25b suggests a dynamical past. While some authors argue that planets in this radius range can be formed \textit{in situ}, these simulations do not generally yield eccentricities as high as that observed for K2-25b \citep{Lee2015BreedingDisks,Ogihara2015ASuper-Earths}. On the other hand, \citet{Moriarty2016THEFORMATION} find disks with shallow surface density profiles produce systems where only one planet will typically transit, where planets have large eccentricities, obliquities, and spacings. In their simulations, the planets in these systems are separated by $20-60$ Hill radii. K2-25b could be an example of such a system; in this case we might expect a second planet with $P\lesssim15$ days that likely does not transit. Our TTV analysis rules out many planets within this period range, but has limited mass sensitivity near $15$ days.

Alternatively, K2-25b could have arrived at its present orbit via a high-eccentricity migration mechanism, which has been proposed as a mechanism of inward migration for hot Jupiters \citep[see][for an overview]{Dawson2018OriginsJupiters}. This process involves two steps: the orbital momentum of the planet is reduced by a perturber, decreasing periastron distance and exciting eccentricity; then its orbital energy is reduced due to tidal dissipation, drawing the planet inward. After this close-in eccentric system is created, its orbit is eventually circularized. 

\subsection{Future outlook}

Further constraints on the presence of a companion require new data. Long-term radial velocity or astrometric monitoring would place additional constraints on the existence of a companion. 
\citet{Mann2016ZODIACALCLUSTER} found no periodic signals at periods $<20$ days in the \emph{K2} data, and no additional transit-like signals were seen in a by-eye inspection.
Measurement of the Rossiter-McLaughlin effect would indicate whether the planetary orbit is aligned with the host star, or if it is misaligned as in the case of Gl 436b \citep{Bourrier2017OrbitalStar}. Significant
spin-orbit misalignment and retrograde motion could result from certain high-eccentricity migration scenarios \citep[e.g.][]{2012ApJ...754L..36N, 2014ApJ...791...86L, 2017MNRAS.465.3927S}.   

\section{Summary}\label{sec:summary}

K2-25b is an eccentric hot Neptune in the Hyades cluster. We obtained multi-year photometric monitoring and $12$ transit lightcurves using the MEarth Observatories. Combing our data with those from \textit{Spitzer}, our dataset consists of 22 non-consecutive transits. The MEarth data are available in a online table. We fit for $\rprstar$ and the transit times $T$ using the \texttt{batman} tool and \texttt{emcee} MCMC package. Our transit time measurements can be found in Table \ref{tab:table2}.

We find that the amplitude of photometric variability was high, at around 2\% peak-to-peak, during the \textit{K2} campaign and the ground-based observing in 2015-2016. From 2016-2018, when most of our transit observations occurred, the variability is significantly diminished; this highlights the importance of long-term photometric monitoring. Despite the number of transits and the presumed overall spottiness of the star, we do not definitively identify spot crossings. There is tentative evidence of a spot crossing in one transit in the {\it Spitzer} data, and two transits from MEarth show systematic deviations from the fiducial transit model. 

The uncertainty in $Q_P$ and the planetary mass, and thus in the circularization timescale $\tau_\mathrm{circ}$, makes it challenging to establish whether K2-25b's eccentricity requires ongoing excitation. However, for Neptunian values of $Q_P$, the age of the planetary system ($650$ Myr) is similar to $\tau_\mathrm{circ}$ ($410$ Myr; 68\% confidence interval: $180$ Myr to $1$ Gyr). 

Our measured transit times show no evidence for TTVs. On this basis, we can exclude roughly Earth-mass companions out to the 2:1 period ratio, and super Earths with masses $>5$ $M_\oplus$ out to the 7:2 perod ratio. Save for in strong resonances, TTVs do not provide strong constraints on companions at longer periods.

\acknowledgments

We thank Jack Defandorf, Debbie Meinbress, and Danielle Noonan at the MIT Kavli Institute for their support, Gongjie Li and Rebekah Dawson for helpful conversations, and the referee for the helpful comments. ERN acknowledges support from the NSF through the Astronomy \& Astrophysics Postdoctoral Fellowship programs under award AST-1602597 and JAD acknowledges support from the Heising-Simons Foundation through the 51 Peg b Fellowship. The MEarth Team gratefully acknowledges funding from the David and Lucille Packard Fellowship for Science and Engineering (awarded to DC). This material is based upon work supported by the National Science Foundation under grants AST-0807690, AST-1109468, AST-1004488 (Alan T. Waterman Award), and AST-1616624. This publication was made possible through the support of a grant from the John Templeton Foundation. The opinions expressed in this publication are those of the authors and do not necessarily reflect the views of the John Templeton Foundation. Support for program GO-14615 was provided by NASA through a grant from the Space Telescope Science Institute, which is operated by the Association of Universities for Research in Astronomy, Inc., under NASA contract NAS 5-26555. This research was supported in part by the National Science Foundation under Grant No. NSF PHY-1748958.
%

\vspace{5mm}
\facilities{K2, Spitzer, MEarth}


\software{astropy \citep{Collaboration2018ThePackage}, matplotlib  \citep{Hunter2007Matplotlib:Environment}, seaborn  \citep{Waskom2018Mwaskom/seaborn:2018}, emcee  \citep{Foreman-Mackey2013EmceeHammer}, batman  \citep{Kreidberg2015Batman:Python}
          }
          
\bibliographystyle{aasjournal}
\bibliography{references_fixed.bib}



\end{document}